\pdfoutput=1
\documentclass[sigconf,nonacm]{acmart}

\usepackage{multirow}
\usepackage{tabularx}
\usepackage{array}
\usepackage{float}

\usepackage{algorithm}
\usepackage{algpseudocode}
\usepackage{xcolor}
\colorlet{BLUE}{blue}
\usepackage{listings}
\usepackage{booktabs,tabularx,array}
\definecolor{CodeBoxBackground}{RGB}{246,246,246}
\definecolor{CodeBoxBorder}{RGB}{180,180,180}

\lstdefinestyle{promptbox}{
    basicstyle=\ttfamily\scriptsize,
    backgroundcolor=\color{CodeBoxBackground},
    frame=single,
    rulecolor=\color{CodeBoxBorder},
    framesep=5pt,
    framexleftmargin=4pt,
    framexrightmargin=4pt,
    linewidth=\linewidth,
    breaklines=true,
    breakatwhitespace=false,
    columns=fullflexible,
    keepspaces=true,
    showstringspaces=false,
    aboveskip=6pt,
    belowskip=8pt
}

\newcolumntype{Y}{>{\raggedright\arraybackslash}X}
\newcolumntype{C}[1]{>{\centering\arraybackslash}m{#1}}
\newcolumntype{L}[1]{>{\raggedright\arraybackslash}m{#1}}
\setcopyright{none}
\renewcommand\footnotetextcopyrightpermission[1]{}

\begin{document}
\raggedbottom
\setlength{\emergencystretch}{2em}

\title{GazeAgent: VLM-Guided Implicit Gaze Calibration from Egocentric Multimodal Behavior for AI Glasses}
\author{Yuhong Zhang}
\affiliation{%
  \department{Department of Bioengineering}
  \institution{University of California San Diego}
  \country{}}

\author{Khanh An Nguyen}
\affiliation{%
  \department{Swartz Center for Computational Neuroscience}
  \institution{University of California San Diego}
  \country{}}

\author{Xinyu Tian}
\affiliation{%
  \institution{Tsinghua University}
  \country{}}
\affiliation{%
  \institution{University of California San Diego}
  \country{}}

\author{KungYi Chang}
\affiliation{%
  \department{Swartz Center for Computational Neuroscience}
  \institution{University of California San Diego}
  \country{}}

\author{Abraham Padua}
\affiliation{%
  \department{Swartz Center for Computational Neuroscience}
  \institution{University of California San Diego}
  \country{}}

\author{Yuchen Xu}
\affiliation{%
  \department{Institute for Neural Computation}
  \institution{University of California San Diego}
  \country{}}

\author{Ying Choon Wu}
\affiliation{%
  \department{Institute for Neural Computation}
  \institution{University of California San Diego}
  \country{}}

\author{Gert Cauwenberghs}
\affiliation{%
  \department{Department of Bioengineering}
  \department{Institute for Neural Computation}
  \institution{University of California San Diego}
  \country{}}

\author{Tzyy-Ping Jung}
\affiliation{%
  \department{Institute for Neural Computation}
  \department{Department of Bioengineering}
  \institution{University of California San Diego}
  \country{}}

\renewcommand{\shortauthors}{Zhang et al.}
\hypersetup{pdfauthor={Yuhong Zhang, Khanh An Nguyen, Xinyu Tian, KungYi Chang, Abraham Padua, Yuchen Xu, Ying Choon Wu, Gert Cauwenberghs, Tzyy-Ping Jung}}

\begin{abstract}
AI glasses typically require explicit, user-specific eye-tracker calibration, and accuracy can drift when the device shifts during use. We present GazeAgent, a VLM-guided framework that continuously and adaptively improves gaze-calibration accuracy from naturally occurring behavior without asking users to interrupt their activities. GazeAgent identifies precise visual attention (PVA) events from multimodal streams of inertial measurement unit (IMU), gaze, and wearer-hand signals, validates their task relevance, and localizes functional targets as pseudo-calibration targets. In a 10-participant scripted study, the system detected all 820 predefined PVA opportunities and, under device disturbance, reduced median gaze error by 44.1\% compared with explicit calibration. In a five-participant case study, autonomous calibration produced the lowest workload and interruption ratings, and all participants preferred it for routine use. By deriving task-critical targets from natural multimodal behavior, GazeAgent demonstrates the potential for continuous, adaptive, and low-burden gaze interaction on egocentric AI glasses.
\end{abstract}



\maketitle

\begin{figure*}[t]
    \centering
    \includegraphics[width=\textwidth]{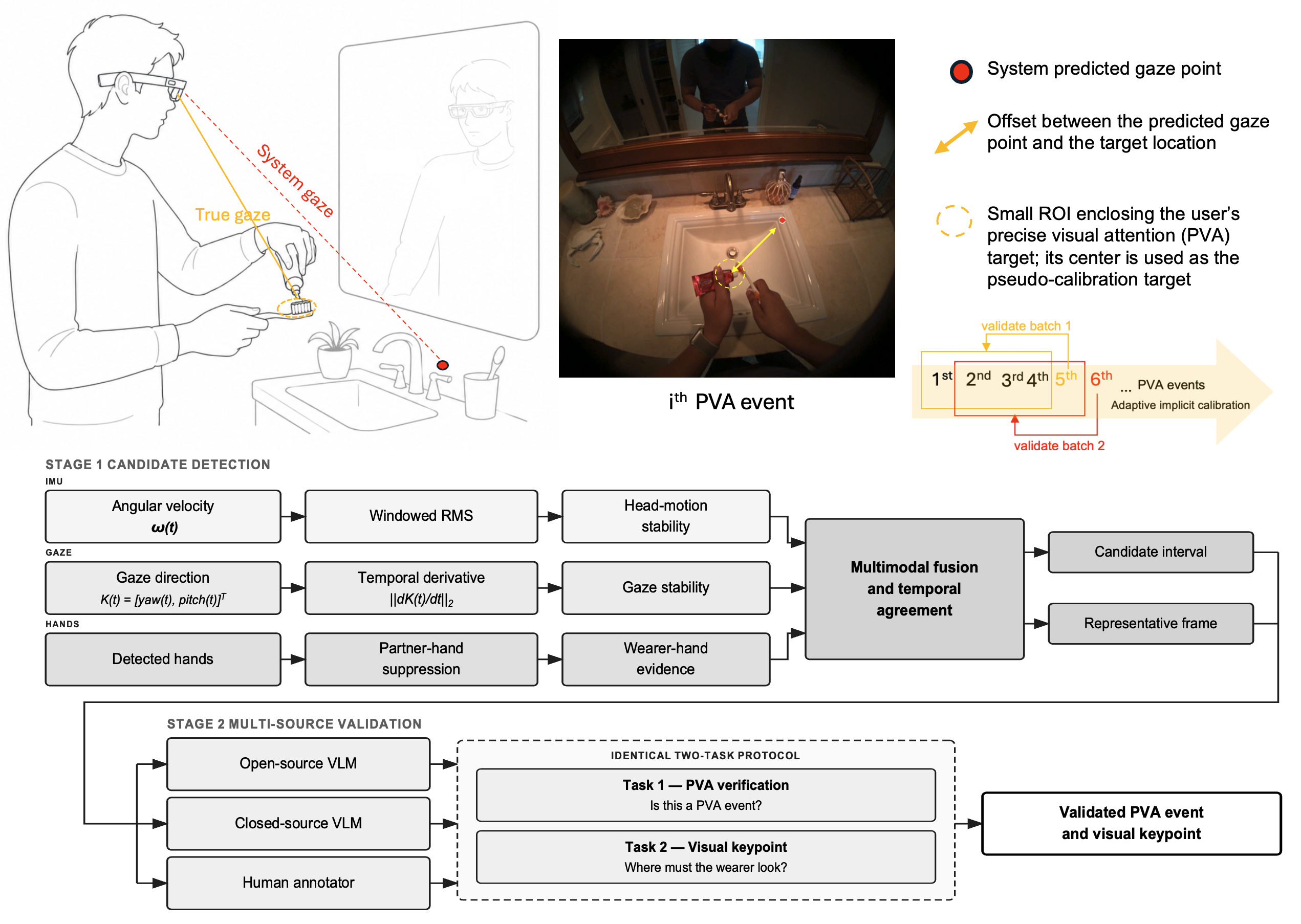}
    \caption{\textbf{Overview of the GazeAgent implicit-calibration pipeline.}
    A qualifying attention event provides a pseudo-calibration target,
    defined as the center of its functional region of interest (ROI), enabling
    estimation of the offset between the system-predicted gaze point and the
    intended target. Stage~1 combines head-motion stability, gaze stability,
    and wearer-hand evidence to generate candidate intervals and representative
    frames. Stage~2 uses human annotators and open-source and closed-source vision-language models (VLMs) to
    validate each candidate and localize its functional-target ROI; the ROI
    center is the single pseudo-calibration keypoint. Validated events are
    processed in successive batches to support continuous, one-step-ahead gaze
    calibration.}
    \Description{Overview of the GazeAgent pipeline. A conceptual example
    illustrates the offset between the predicted gaze point and an event-based
    pseudo-calibration target. A multimodal candidate detector combines IMU,
    gaze, and hand signals. Human annotators and vision-language models then
    validate candidate events and localize their functional-target ROIs and
    center keypoints. Successive batches support adaptive implicit calibration.}
    \label{fig:gazeagent-overview}
\end{figure*}

\section{Introduction}\label{sec:introduction}

AI glasses have the potential to become a general-purpose wearable computing platform~\cite{engel2023projectaria,plopski2022eye}. Their egocentric cameras and multimodal sensors provide a continuous view of the wearer's activities, while eye tracking offers a direct signal of visual attention~\cite{engel2023projectaria,min2021integrating}. Reliable gaze estimation can support activity recognition and context-aware assistance~\cite{engel2023projectaria,min2021integrating}. It also enables hands-free target selection and other gaze-based interactions with accessibility potential~\cite{wei2023predicting,plopski2022eye}. Accurate and continuously available gaze estimates are therefore important for understanding user attention through in-the-wild egocentric AI glasses~\cite{engel2023projectaria,plopski2022eye}.

Current head-mounted eye trackers predominantly rely on explicit, user-specific calibration. When initializing an augmented- or extended-reality (AR/XR) headset or AI glasses system, users are typically instructed to fixate on a sequence of predefined targets distributed across the field of view~\cite{clay2019eye,liu2024personal}. Although pursuit-based procedures can make calibration less tedious and more flexible, they still depend on users following known moving targets rather than deriving references from unconstrained ongoing activities~\cite{pfeuffer2013pursuit}.

Moreover, an initially accurate calibration may drift during real-world use. A user may remove and put on the glasses again, adjust the frame for comfort, or shift the device during motion-intensive activities such as walking, jogging, or jumping. These changes alter the relative positions of the eye cameras and the wearer's eyes, introducing systematic drift into the estimated gaze direction. Consequently, gaze accuracy may deteriorate during a session even when the initial explicit calibration was performed correctly.

Implicit calibration aims to address this problem by deriving calibration correspondences from naturally occurring behavior rather than explicitly presented targets~\cite{sugano2015self,chen2014probabilistic,10.1109/INFOCOM42981.2021.9488668}. A known point in the scene can stand in for an explicitly presented calibration marker. When a user taps a button in a graphical interface, the tapped element is almost certainly where they were looking, so its image coordinates and the gaze coordinates recorded at the same timestamp form one gaze-target pair \cite{10.1145/3706598.3713461,10.1109/INFOCOM42981.2021.9488668}. A handful of such pairs is enough to fit a low-parameter transformation, in our case a six-coefficient affine mapping, that realigns the raw predicted gaze coordinates with the intended targets and absorbs the offset and scale error introduced by device slippage and drift \cite{10.1145/2807442.2807445,sugano2015self}. The correction applies to the output of the gaze estimator rather than its input. The mapping from eye-camera images to gaze vectors is left unchanged; only the residual drift of the existing estimate is removed. Figure~\ref{fig:gazeagent-overview} shows the offset between a predicted gaze point and such a target. However, existing methods commonly rely on constrained environments in which the target is already known, such as a user interface element, a word on a screen, or another predefined point of interest. Comparable targets are difficult to identify in unconstrained egocentric scenes, where multiple objects may be visible and visual salience does not necessarily indicate the functional target of an action. Earlier approaches based on conventional saliency estimation or screen-based useful-frame selection~\cite{10.1145/2807442.2807445,10.1109/INFOCOM42981.2021.9488668} therefore do not directly address continuous calibration for in-the-wild egocentric AI glasses.

In this study, we define a PVA event as a temporally bounded interval during which successful task completion requires the wearer to foveate a spatially precise, task-relevant functional target, such as a contact point, insertion point, alignment feature, or marked location. The functional target must occupy a relatively small portion of the full field of view, requiring localized central vision. Without direct foveation of that region, the task would be likely to fail or become error-prone. In contrast, coarse reaching or grasping actions that do not require precise visual guidance are not considered PVA events. Table~\ref{tab:pva_examples} contrasts accepted and rejected examples. For each accepted event, the model predicts a functional-target ROI and its center serves as the single pseudo-calibration keypoint.

This formulation differs from general egocentric gaze or saliency prediction, which estimates visually likely regions from scene context~\cite{park2026robustegocentricvisualattention,lai2024listenlookfutureaudiovisual,huang2018predictinggazeegocentricvideo,huang2020an}. Our objective is instead to identify discrete PVA intervals and localize the functional ROI associated with each event. For example, when inserting a connector, the relevant target is the receiving port rather than the hand, the connector body, or the surrounding object.

\begin{table*}[t]
\centering
\caption{
Representative accepted and rejected examples of PVA
events. Accepted events provide localized pseudo-calibration targets,
whereas rejected events do not require precise foveal attention.
}
\label{tab:pva_examples}

\setlength{\tabcolsep}{4pt}
\renewcommand{\arraystretch}{1.05}
\footnotesize

\begin{tabularx}{\textwidth}{YYYY}
\toprule
\multicolumn{2}{c}{\textbf{Accepted Events}} &
\multicolumn{2}{c}{\textbf{Rejected Events}} \\
\cmidrule(lr){1-2}
\cmidrule(lr){3-4}

\includegraphics[
    width=\linewidth,
    height=0.14\textheight,
    keepaspectratio
]{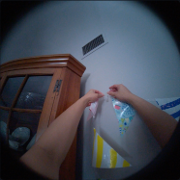}
&
\includegraphics[
    width=\linewidth,
    height=0.14\textheight,
    keepaspectratio
]{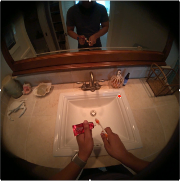}
&
\includegraphics[
    width=\linewidth,
    height=0.14\textheight,
    keepaspectratio
]{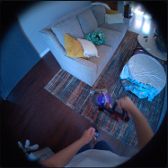}
&
\includegraphics[
    width=\linewidth,
    height=0.14\textheight,
    keepaspectratio
]{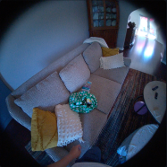}
\\[-1mm]

\centering\textbf{Peeling a tape strip}
&
\centering\textbf{Applying toothpaste}
&
\centering\textbf{Reaching for a large object}
&
\centering\textbf{Reaching for a pillow}
\tabularnewline

\centering\textbf{Accept}
&
\centering\textbf{Accept}
&
\centering\textbf{Reject}
&
\centering\textbf{Reject}
\tabularnewline

\textbf{Target:} Tape edge
&
\textbf{Target:} Toothbrush tip
&
\textbf{Target:} None
&
\textbf{Target:} None
\\

\textbf{H1:} The tape edge is a spatially precise functional target.

\textbf{H2:} The head and hands remain stable.

\textbf{H3:} Both hands perform a fine peeling action.

\textbf{H4:} The action involves sustained manipulation and an elevated
reach.
&
\textbf{H1:} Toothpaste must be accurately placed on the toothbrush head.

\textbf{H2:} The head, hands, tube, and toothbrush remain stable.

\textbf{H3:} One hand stabilizes the toothbrush while the other squeezes
the tube.

\textbf{H4:} The action requires sustained attention and precise
alignment.
&
\textbf{H1:} No spatially precise functional target is visible.

\textbf{H2:} The user appears to be moving toward the object.

\textbf{H3:} The hand performs a coarse reaching action.

\textbf{H4:} The action is brief, familiar, and visually undemanding.
&
\textbf{H1:} The large target does not require precise foveal vision.

\textbf{H2:} The frame shows repositioning rather than stable fixation.

\textbf{H3:} The hand performs a coarse grasp.

\textbf{H4:} The action is familiar and low precision.
\\

\bottomrule
\end{tabularx}
\end{table*}

We present GazeAgent, a VLM-guided framework that derives task-critical calibration targets from natural multimodal behavior rather than bottom-up saliency or predefined interaction coordinates. A high-recall multimodal behavioral detector identifies candidate PVA events from multimodal agreement among head-motion stability, raw-gaze stability, and wearer-hand presence. A fine-tuned open-source VLM validates each event and predicts its functional-target ROI, whose center defines the pseudo-calibration target. Human annotations provide evaluation labels and fine-tuning supervision, and successive batches of validated gaze-target correspondences update an affine calibration model.

{PVA event selection and validation are organized around four hypotheses. \textbf{H1: Task necessity and functional precision} asks whether successful task completion requires direct attention to a functional target that is relatively small within the full field of view. \textbf{H2: Motion stability} considers whether head and gaze stabilization marks a usable fixation interval. \textbf{H3: Hand-eye consistency} examines whether gaze converges on the object or feature involved in fine hand action. \textbf{H4: Temporal and physical context} has two components: \textbf{H4.1: Sustained attention} considers whether PVA persists across the task-relevant interval, and \textbf{H4.2: Reachability demand} considers whether difficult-to-reach targets increase the need for close visual guidance. Section~\ref{sec:hypothesis-validation} evaluates evidence independent of candidate selection and reports detector-conditioned signals descriptively; the VLM reasons jointly over all four hypotheses rather than applying a fixed logical rule.}


Our results show that the fine-tuned open-source VLM reliably validates candidate PVA events and localizes their functional targets. The resulting implicit calibration reduces angular gaze error relative to the pre-trained model and, in our evaluation, the default explicit procedure. Naturally occurring PVA events can therefore provide useful pseudo-calibration references without repeatedly interrupting the wearer.

In summary, this work makes the following contributions:

\begin{itemize}\item {To the best of our knowledge, we introduce the first VLM-guided implicit-calibration framework for egocentric AI glasses that uses naturally occurring PVA events as sources of task-critical functional targets.}

\item We develop a high-recall multimodal detector that combines head-motion stability, gaze stability, and wearer-hand evidence to propose candidate PVA intervals efficiently.

\item We present a two-task VLM framework that validates candidate PVA events and localizes their functional-target ROIs and center keypoints using temporal context, multimodal evidence, and human supervision.

\item We develop an adaptive implicit-calibration method that updates an affine gaze transformation from consecutive PVA-based pseudo-calibration points.

\item We evaluate the complete pipeline across everyday activities and controlled disturbance conditions, demonstrating improved gaze accuracy without requiring users to interrupt their activities for repeated explicit calibration.\end{itemize}

\section{Related Work}
\label{sec:related}

\subsection{Implicit Calibration for Head-Mounted Eye Tracking}
\label{sec:related-implicit-calibration}

Explicit target-based calibration remains the predominant approach for
head-mounted eye tracking \cite{clay2019eye,liu2024personal}. Although
it can provide accurate initial estimates, calibration interrupts the
user and may become inaccurate when the device shifts or the eye-camera
geometry changes. Implicit methods instead derive supervision from the
user's ongoing visual experience. Sugano and Bulling use egocentric
visual saliency to update a head-mounted gaze model
\cite{sugano2015self}, while Chen and Ji probabilistically relate eye
appearance to the saliency distribution of displayed content
\cite{chen2014probabilistic}. vGaze similarly uses saliency-aware frame
selection for continuous mobile gaze calibration
\cite{10.1109/INFOCOM42981.2021.9488668}. These methods avoid dedicated
calibration targets, but assume that visually salient regions correspond
to the user's intended fixation targets.

Other methods obtain pseudo-calibration targets from interaction events.
Online-EYE uses the known locations of UI elements selected through
controllers or direct touch to continuously update eye tracker
calibration in XR \cite{10.1145/3706598.3713461}. This provides stronger
supervision than saliency because the interface supplies the selected
target's coordinates. However, it requires an instrumented environment
in which interaction targets and their locations are known. In physical
environments, the relevant location may instead be a button, opening,
edge, or contact point whose coordinates are not predefined. Existing
implicit calibration approaches therefore rely primarily on either
bottom-up saliency or explicit interaction coordinates; identifying
task-relevant calibration targets in unstructured egocentric scenes
remains less explored.

\subsection{Behavioral Foundations of Pseudo-Calibration Events}
\label{sec:related-pva-hypotheses}

Research on visually guided action provides evidence for task necessity,
foveal demand, and hand-eye consistency. Pointing accuracy decreases
when the target is not directly fixated
\cite{neggers2000gaze}, and gaze is directed toward task-relevant contact
points and landmarks during object manipulation
\cite{johansson2001eye}. Gaze also commonly reaches a target before the
hand and remains near the relevant region during grasping, transport,
and placement
\cite{johansson2001eye,flanagan2003action,land1999roles,
10.1098/rstb.2021.0453}. These findings already partially support H1 and H3: a useful
pseudo-calibration event should involve a visually necessary target and
show spatial or temporal consistency between gaze and manual action.
They also suggest that target size should be interpreted functionally,
because a relatively large object may contain a small contact point that
requires precise central vision.

Motion stability and event duration provide additional evidence about
whether a candidate contains a usable fixation interval. Low head motion
may support stable gaze, although compensatory eye movements can
maintain fixation while the head moves
\cite{land2006eye,kothari2020gaze,drews2024strategies}. Head stability is
therefore a supporting signal rather than a necessary condition.
Fixation-detection methods generally combine spatial stability with a
minimum duration \cite{salvucci2000identifying}, while natural actions
may contain multiple predictive and feedback-driven fixations
\cite{land1999roles,land2006eye}. Interaction-based calibration also
restricts samples according to their duration and temporal proximity to
an action \cite{10.1145/3706598.3713461}. These findings motivate H2 and
H4, concerning head-gaze stability and sustained attention. Physical
reach effort may affect visual sampling
\cite{matthis2018gaze,bonnen2021binocular}, but does not by itself show
that a particular location must be fixated.

\subsection{VLMs for Egocentric and Gaze-Aware Video Understanding}
\label{sec:related-egocentric-vlms}

VLMs have been extended from static images to video through temporal
feature aggregation, longer context windows, and more efficient visual
representations. Egocentric benchmarks evaluate whether these models can
reason about activities from the camera wearer's perspective. EgoTaskQA
tests spatial, temporal, causal, and goal-oriented reasoning
\cite{jia2022egotaskqa}; EgoSchema requires evidence distributed across
extended first-person videos \cite{mangalam2023egoschema}; and EgoThink
evaluates first-person reasoning \cite{cheng2024egothink}. EgoVLPv2
learns video-language representations adapted to egocentric activities
\cite{pramanick2023egovlpv2}, while ECBench and EgoTextVQA evaluate
embodied cognition and scene-text reasoning
\cite{dang2025ecbench,zhou2025egotextvqa}. These studies demonstrate
progress in activity and goal understanding but also identify continuing
difficulties in fine-grained temporal reasoning and precise spatial
grounding.

Most egocentric VLMs primarily use RGB and language, with gaze and other
wearable sensor signals receiving less attention. Gaze-VLM uses recorded
human gaze to regularize model attention during training but does not
require gaze during inference \cite{pani2025gazevlm}. EgoGazeVQA instead
provides gaze during inference and evaluates its contribution to
spatial, temporal, and intent reasoning
\cite{peng2025eyemllm}. The distinction is important: training-time gaze
acts as supervision, whereas inference-time gaze remains dependent on
the accuracy of the wearable eye tracker. Both approaches use
pre-recorded gaze from existing datasets and do not address changes in
gaze accuracy caused by device movement or calibration drift during
continued use. Thus, gaze-aware egocentric reasoning and long-term
maintenance of gaze calibration remain largely separate research
problems.

\section{GazeAgent}\label{sec:gazeagent}
\subsection{Motivation}\label{sec:gazeagent-motivation}
Implicit calibration can restore gaze accuracy after device shifts without requiring users to repeat an explicit calibration procedure, but unconstrained egocentric activities provide no predefined targets from which to derive calibration correspondences. GazeAgent addresses this requirement by combining multimodal event proposals, VLM-based semantic validation, functional-target localization, and confidence-aware calibration updates.

Prior agentic frameworks have shown that language-model-based agents can interleave reasoning and action, invoke external tools, and incorporate human input into multi-step workflows
\cite{yao2023react,wu2024autogen}.
GazeAgent's core calibration loop is autonomous: uncertain candidates are rejected or deferred rather than requiring a response. An optional personalization mode can query users at natural pauses and use their confirmations or corrections for subject-specific adaptation.

\begin{figure*}[t]
    \centering
    \includegraphics[width=\textwidth]{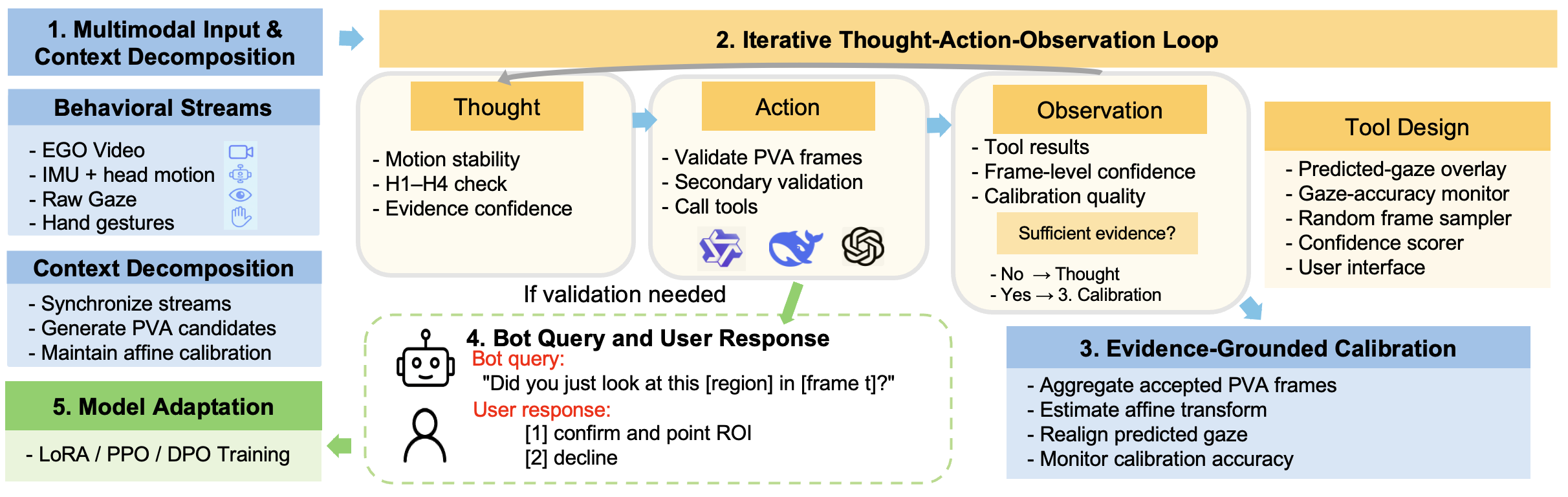}
    \caption{Overview of the GazeAgent workflow, which integrates multimodal behavioral streams, iterative VLM reasoning and tool use, evidence-grounded implicit calibration, and optional feedback-driven personalization.}
    \Description{GazeAgent synchronizes egocentric video, motion, gaze, and hand signals; validates candidate PVA events through an iterative thought, action, and observation loop; performs evidence-grounded calibration; and can optionally collect feedback for personalization.}
    \label{fig:gazeagent-workflow}
\end{figure*}

\subsection{Agent Workflow}\label{sec:agent-workflow}

As shown in Figure~\ref{fig:gazeagent-workflow}, GazeAgent operates in the background on synchronized streams of egocentric video, IMU and head-motion measurements, raw-gaze, and wearer-hand activity. It decomposes these inputs to identify candidate PVA intervals and their representative frames while maintaining the current affine transformation for gaze realignment.

For each candidate, GazeAgent executes an iterative thought, action, and observation loop. In the \textit{thought} stage, the agent evaluates motion stability, evidence associated with {the four hypotheses (H1 to H4, with H4.1 and H4.2 as components of H4)}, and its overall confidence. In the \textit{action} stage, it may accept or reject the candidate, invoke a validation tool, or request additional evidence. In the \textit{observation} stage, the agent incorporates tool outputs, frame-level confidence, and changes in calibration quality into its context. This process continues until sufficient evidence is available to make a decision.

Accepted candidates are collected as validated gaze-target correspondences, denoted by $\mathcal{V}_t$, and used to update the affine transformation:
\begin{equation}
    \mathbf{A}_{t+1}
    =
    \mathrm{FitAffine}\!\left(\mathcal{V}_t\right).
    \label{eq:affine-update}
\end{equation}
The updated transformation is then applied to subsequent gaze predictions. In the autonomous calibration mode, uncertain candidates are excluded. The optional personalization mode may instead present selected low-confidence cases at natural pauses, allowing the user to confirm the prediction, correct the ROI, or reject the candidate. The main VLM is adapted through supervised fine-tuning (SFT); the subject-wise case study in Section~\ref{sec:case-study-adaptation} uses Direct Preference Optimization (DPO) on optional user feedback.

\subsection{Tool Design}
\label{sec:tool-design}

GazeAgent uses five tools that provide structured evidence to the thought, action, and observation loop. \textbf{(1) Predicted-gaze overlay} is a post-hoc validation and sanity-check visualization. Absolute gaze position and gaze overlays are not supplied to the VLM when it validates an event or localizes its pseudo-ground-truth target, preventing circular target selection. \textbf{(2) Gaze-accuracy monitor} tracks performance after validated references become available. \textbf{(3) Random frame sampler} audits selection bias across times, activities, and confidence levels. \textbf{(4) Confidence scorer} combines motion stability, VLM consistency, sensor quality, and calibration trends to accept, reject, or defer candidates. \textbf{(5) User interface} is enabled only for optional personalization and allows a user to confirm a prediction, correct the ROI, or reject a target. Section~\ref{sec:case-study-interface} presents the detailed interface design and its use in the personalization case study.

\section{Candidate PVA Event Detection from Generic Multimodal Behavioral Streams}
\label{sec:candidate-detection}

\textbf{RQ1: Can generic multimodal behavioral signals be used to identify PVA events?}
To answer this question, we describe the datasets and candidate-detection algorithm, then use human annotations to characterize the detected events. The following section evaluates whether VLMs can reproduce these human judgments and localize their functional targets.

We use two datasets in this study. First, the publicly available Aria Everyday Activities (AEA) dataset~\cite{lv2024aria} provides diverse, in-the-wild recordings for evaluating multimodal candidate detection and subsequent VLM-based event validation. The multimodal detector initially identifies 1,212 candidate events across 143 recordings. These candidates are proposals rather than confirmed PVA events and may therefore include false positives. We aggregate annotations from multiple human raters into consensus labels, which serve as the reference for evaluating the VLM. The resulting human-labeled examples are also used for parameter-efficient supervised fine-tuning with Low-Rank Adaptation (LoRA).

Second, we collect a dataset to evaluate the continuous adaptation of our implicit-calibration algorithm. The AEA dataset contains a median of six candidate events per recording, and downstream validation rejects some proposals. Because many AEA recordings therefore contain too few reliable targets for repeated updates, {we recruited 10 participants, each of whom completed both scripted task routes, to elicit sustained sequences of PVA events; Section~\ref{sec:experimental-protocol} presents the detailed protocol.}

\subsection{Aria Everyday Activities Dataset}
\label{sec:aea-dataset}

AEA is a publicly available multimodal egocentric dataset captured using Meta Project Aria glasses. It comprises 143 recordings totaling more than 7.5 hours and one million images across five indoor locations, with participants performing everyday activities such as cooking, exercising, playing games, and socializing. In addition to raw sensor streams, AEA provides globally aligned 3D device trajectories, semi-dense scene point clouds, per-frame 3D gaze estimates, online camera and IMU calibration, and time-aligned speech transcriptions. The synchronized egocentric video, gaze, and motion signals make AEA well suited to evaluating GazeAgent under realistic conditions, including natural head movement, changes in wearer position, hand-object interactions, and diverse scene contexts.

\subsection{Multimodal Agreement for PVA Candidate Detection}\label{sec:multimodal-detector}

\begin{figure*}[t]\centering\includegraphics[width=\textwidth]{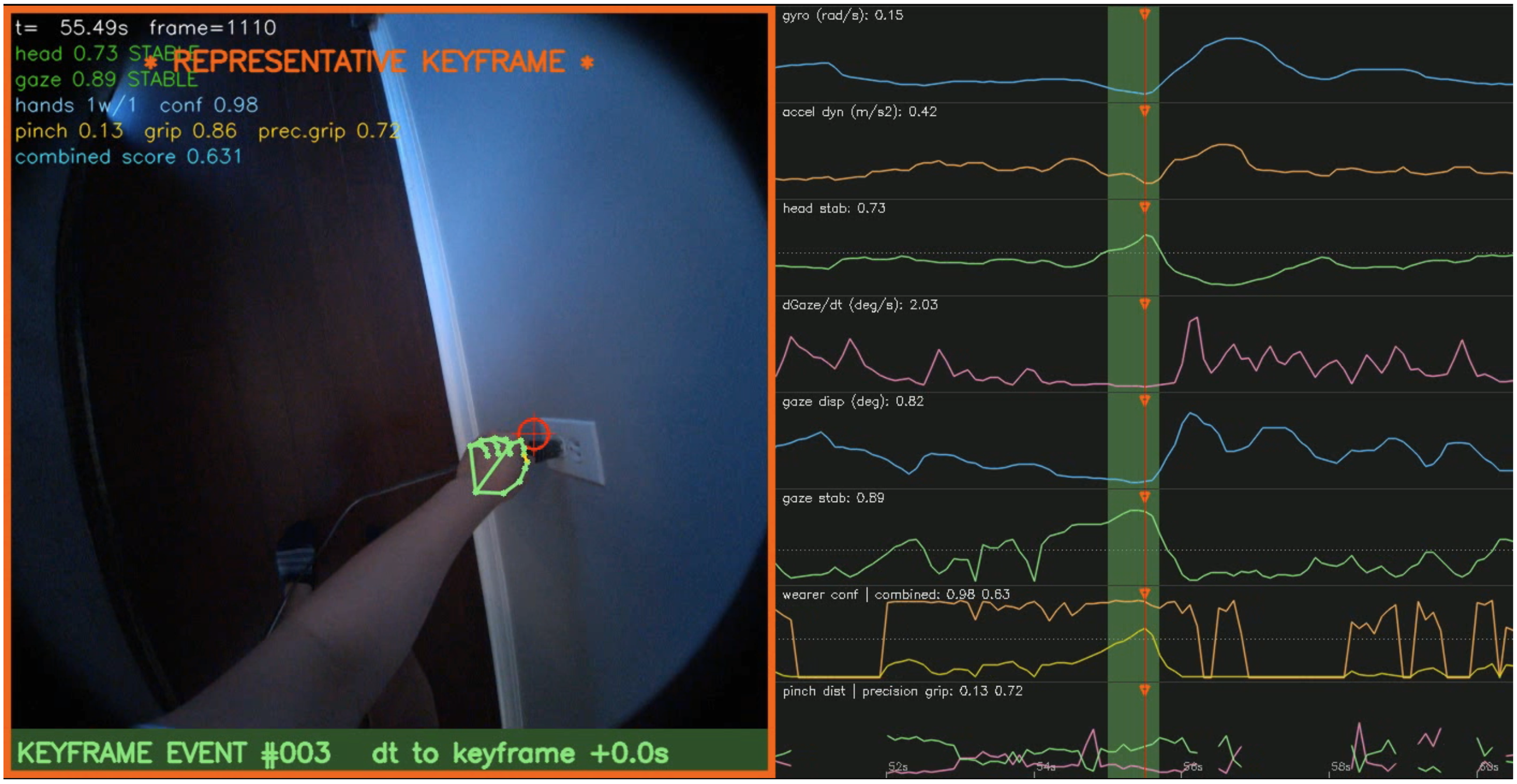}\caption{\textbf{Example output of the high-recall multimodal PVA candidate detector.} The left panel shows the representative keyframe, wearer-hand overlay, modality scores, and combined ranking score. The displayed gaze marker is used only for post-hoc validation and is removed from every frame supplied to the VLM. The right panel shows the synchronized head-motion, gaze-dynamics, and hand-activity signals. The green region denotes the candidate interval, and the orange line marks its highest-scoring keyframe. The detector proposes a possible event; the VLM subsequently determines its semantic validity without access to absolute gaze position.}\Description{A representative egocentric keyframe is displayed beside synchronized plots of head motion, gaze stability, and hand-interaction signals. A green vertical band indicates the detected candidate interval, while an orange line identifies the selected representative keyframe.}\label{fig:multimodal-stream}\end{figure*}

We design a high-recall multimodal behavioral detector rather than a semantic PVA classifier. It identifies temporal agreement among three behavioral cues: head stability, gaze stability, and wearer-hand presence. Their co-occurrence identifies intervals in which a PVA event may occur without recognizing the underlying object or action. This design intentionally permits false-positive proposals because downstream VLM validation can reject them, whereas a missed event cannot be recovered. Figure~\ref{fig:multimodal-stream} illustrates the synchronized signals, candidate interval, and representative keyframe.

We align the RGB, gaze, and IMU streams to a common 10-Hz analysis timeline. We use $i$ to index the samples on this timeline and $t_i$ to denote the timestamp of sample $i$. Head stability at sample $i$, denoted by $S_i^{H}\in[0,1]$, is estimated from acceleration variation and root-mean-square angular velocity over a centered 1.0-s window, where the superscript $H$ denotes head motion and a larger value indicates greater stability. Gaze stability, denoted by $S_i^{G}\in[0,1]$, combines angular gaze velocity and gaze dispersion over a centered 0.8-s window, where the superscript $G$ denotes gaze and a larger value again indicates greater stability. Because these gaze features describe temporal variation rather than absolute gaze position, the detector remains insensitive to a locally constant calibration offset.

For each RGB frame at sample $i$, we additionally record the number of detected wearer hands, $N_i$, and the confidence of the strongest wearer-hand detection, $C_i\in[0,1]$. We set $C_i=0$ when no wearer hand is detected. The binary multimodal-agreement decision $z_i$ is defined as\begin{equation}z_i =\mathbb{I}\!\left[S_i^{H} \geq 0.5\land S_i^{G} \geq 0.4\land N_i \geq 1\land C_i \geq 0.5\right],\label{eq:multimodal-agreement}\end{equation}where $\mathbb{I}[\cdot]$ is the indicator function: $z_i=1$ when all four conditions are satisfied and $z_i=0$ otherwise. The thresholds require a head-stability score of at least $0.5$, a gaze-stability score of at least $0.4$, at least one visible wearer hand, and a hand-detection confidence of at least $0.5$. These threshold values were selected empirically based on the best observed detection performance. These operating thresholds favor recall over precision. Consequently, some selected frames may contain stable gaze and hand activity without a genuine PVA requirement, but the detector is less likely to discard true PVA events before semantic validation. {Because the detector selects candidates using head stability, gaze stability, gaze dispersion, and wearer-hand presence, selected events will naturally exhibit these properties. We therefore report these measures as characteristics of detector-selected events rather than as independent evidence supporting H1 to H3.}

To form temporally coherent candidate intervals, we median-filter the binary sequence ${z_i}$ over 0.5~s, bridge inactive gaps of at most 0.7~s, discard active runs shorter than 0.6~s, and merge intervals separated by less than 2.0~s. We use $e$ to index the resulting candidate intervals and denote the $e$-th interval by $E_e$. Each interval $E_e$ therefore contains a temporally contiguous set of samples that satisfy the multimodal-agreement criteria after temporal filtering.

Within each candidate interval, we assign every eligible sample a keyframe-quality score $Q_i$:\begin{equation}\begin{aligned}Q_i &=\left(S_i^{H}+\epsilon\right)\left(S_i^{G}+\epsilon\right)\left(C_i+\epsilon\right), \\ i_e^{*} &=\underset{i:\,t_i\in E_e,\,z_i=1}{\arg\max}\; Q_i .\end{aligned}\label{eq:keyframe-score}\end{equation}Here, $Q_i$ is the multiplicative agreement score for sample $i$, and $\epsilon>0$ prevents a zero-valued component from collapsing the product. The detector outputs $E_e$, its representative keyframe $i_e^{*}$, and the component scores $S_{i_e^{*}}^{H}$, $S_{i_e^{*}}^{G}$, and $C_{i_e^{*}}$. Appendix~\ref{app:candidate-detector} presents the complete candidate-detection algorithm.

\begin{figure*}[!t]
    \centering
    \includegraphics[width=\textwidth]{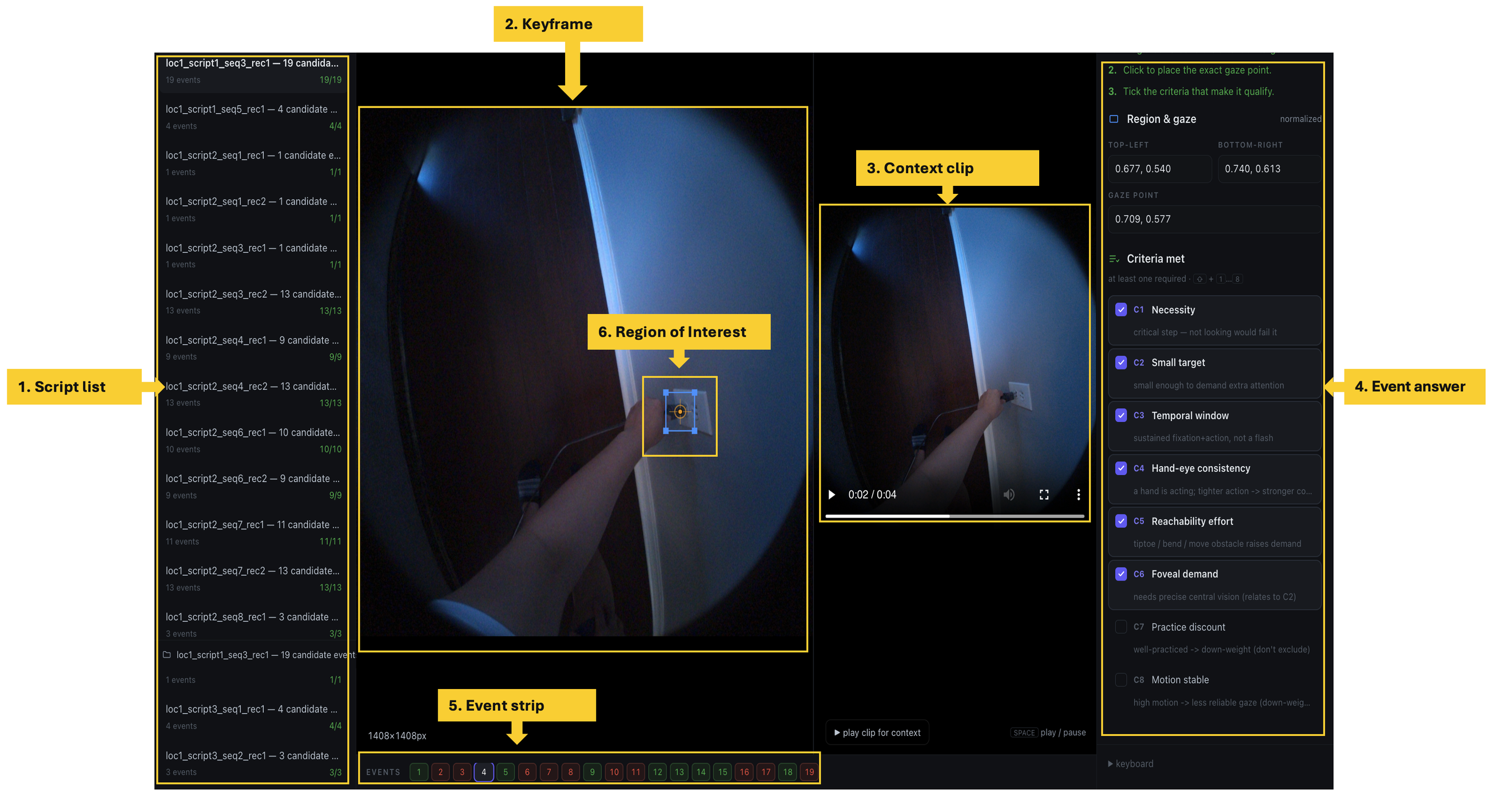}
    \caption{\textbf{Human annotation interface.} Annotators selected a candidate from the script list and event strip, inspected its keyframe and context clip, drew a tight ROI around the functional target, and completed the criterion-level judgments. The point shown inside the ROI is its computed center and serves as the single reference keypoint.}
    \Description{A screenshot of the annotation interface with labeled regions for the script list, keyframe, context clip, event-answer panel, event strip, and functional-target ROI.}
    \label{fig:annotation-interface}
\end{figure*}

\subsection{Human Annotation Protocol and Interface}\label{sec:annotation-protocol}

Five university student annotators independently reviewed the 1,212 candidate events extracted from 143 AEA recordings. For each candidate, they determined whether it was a valid PVA event, identified its functional visual target, and drew a tight ROI around that target. The reference keypoint was computed as the ROI center. They also provided criterion-level judgments corresponding to H1 to H4. Unanimous acceptance by all five annotators defined the human-consensus PVA label used for model evaluation.

Figure~\ref{fig:annotation-interface} shows the event-review workflow. The script list organized recordings and annotation progress, and the event strip selected a candidate keyframe. The context clip allowed annotators to review the video segment containing the candidate event when a single image did not provide enough context for a reliable judgment. Annotators used the event-answer panel to record the overall PVA judgment and criterion-level evidence, then drew a tight ROI around the functional target. The interface displayed the derived ROI-center keypoint for verification, and all spatial annotations were stored as normalized coordinates.

The event-answer panel operationalizes eight criteria: necessity (C1), functional-target scale (C2), temporal window (C3), hand-eye consistency (C4), reachability effort (C5), foveal demand (C6), practice discount (C7), and motion stability (C8). Necessity (C1) asks whether the action would fail or become substantially more error-prone without precise fixation on the functional target, as when inserting a key into a lock rather than grasping a mug. C2 requires the functional target to be relatively small within the full field of view rather than imposing one universal angular-size cutoff. Annotators considered the complete evidence profile when assigning the overall label; criterion selections preserve the reasons for that judgment. Appendix~\ref{app:task1-prompt} provides the full operational definitions.

\begin{figure*}[!t]\centering\includegraphics[width=\textwidth]{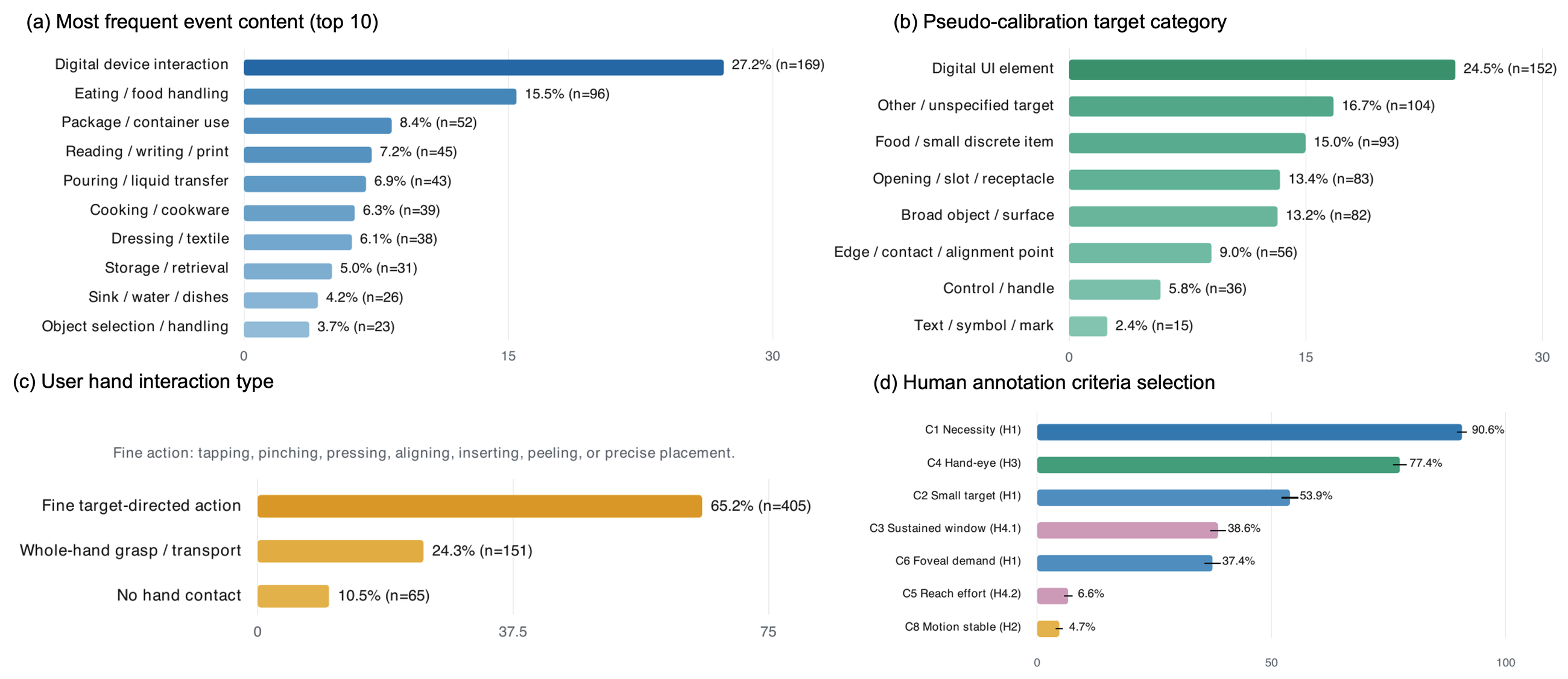}\caption{\textbf{Content and annotation characteristics of human-consensus PVA events.} Results are reported for 621 candidates accepted by all five annotators. \textbf{(a)} Most frequent activity categories. \textbf{(b)} Functional target categories used as pseudo-calibration references. \textbf{(c)} Wearer-hand interaction types. Fine target-directed actions include tapping, pinching, pressing, aligning, inserting, peeling, and precise placement. \textbf{(d)} Annotation-criterion selection rates among accepted judgments. Activity, target, and interaction categories are mutually exclusive within their respective panels; multiple annotation criteria could be selected for the same event.}\Description{Four horizontal bar charts summarize the activity categories, functional targets, hand-interaction types, and annotation criteria of 621 human-consensus PVA events.}\label{fig:pva-event-statistics}\end{figure*}

\subsection{PVA Candidate-Event Statistics after Human Annotation}\label{sec:pva-candidate-statistics}

Figure~\ref{fig:pva-event-statistics} summarizes the 621 events accepted by all five annotators. Digital-device interaction, primarily phone-tapping interactions performed near the beginning and end of each recording, was the largest activity category (27.2\%), but accepted events also spanned eating, reading, pouring, cooking, dressing, and package handling. Functional targets were diverse: digital interface elements accounted for 24.5\%, followed by food or small discrete items, openings or receptacles, broad surfaces, alignment points, controls, and printed marks. Fine target-directed actions accounted for 65.2\%, while whole-hand actions (24.3\%) and events without hand contact (10.5\%) show that PVA is not synonymous with fine manipulation. Annotators selected task necessity (90.6\%) and hand-eye consistency (77.4\%) most often. Motion stability was enforced empirically by the Stage~1 IMU and gaze gates for every proposal, while the remaining criteria described why each stable interval did or did not require PVA. Taken together, these statistics indicate that human-consensus PVA events are neither rare nor confined to a narrow class of fine manipulation: they arise across a broad range of everyday activities, target types, and hand-interaction styles, including the 34.8\% of accepted events that involve whole-hand actions or no hand contact at all. Because task necessity and hand-eye consistency, rather than visual salience, are the criteria that most often justify acceptance, everyday behavior appears to supply functional targets that are both frequent enough and diverse enough to serve as pseudo-calibration references throughout ordinary daily activity.

\begin{figure*}[t]
    \centering
    \includegraphics[width=\textwidth]
    {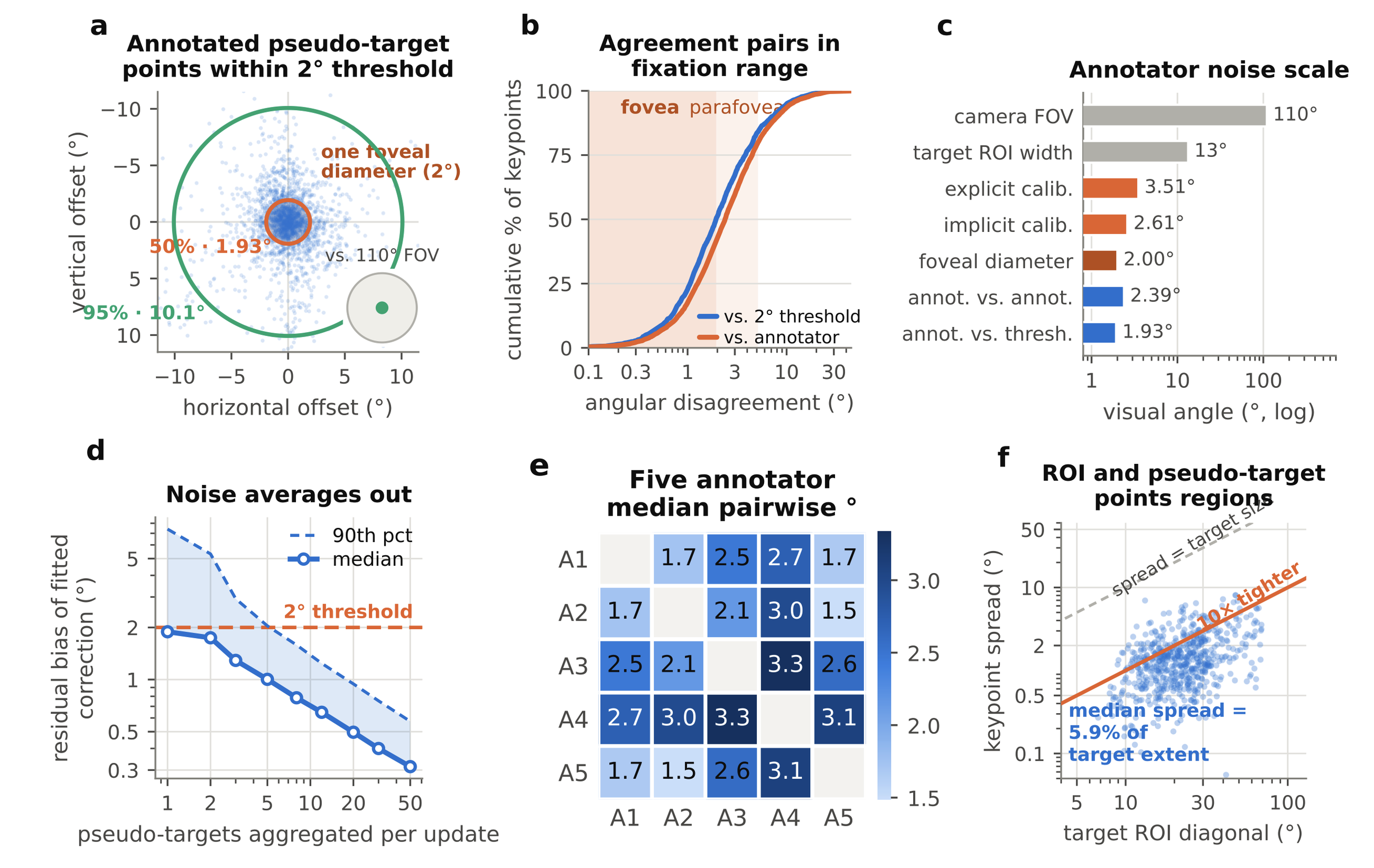}
    \caption{\textbf{Spatial reliability of the human-annotated pseudo-targets.}
    Five annotators independently marked the functional target in each
    accepted PVA event.
    \textbf{(a)} Angular offsets from the leave-one-out annotator consensus.
    The median offset was \(1.93^{\circ}\), approximately one foveal diameter,
    while the 95th-percentile offset was \(10.1^{\circ}\).
    \textbf{(b)} Cumulative distributions of annotator-to-consensus and
    pairwise annotator disagreement; median pairwise disagreement was
    \(2.39^{\circ}\).
    \textbf{(c)} Annotation disagreement relative to the foveal diameter,
    target width, camera field of view, and calibration-error scales.
    \textbf{(d)} Residual correction bias decreases as more pseudo-targets
    are aggregated, from \(1.89^{\circ}\) for one target to \(0.78^{\circ}\)
    for eight and \(0.50^{\circ}\) for twenty.
    \textbf{(e)} Median disagreement between each annotator pair ranged from
    \(1.5^{\circ}\) to \(3.3^{\circ}\).
    \textbf{(f)} Annotated points were substantially more concentrated than
    their corresponding ROIs: the median point spread was 5.9\% of the
    target extent.}
    \Description{Six panels quantify agreement among five annotators,
    compare annotation error with relevant visual-angle scales, and show
    how aggregating pseudo-targets reduces residual correction bias.}
    \label{fig:pseudotarget-reliability}
\end{figure*}

To assess pseudo-target reliability independently of the VLM predictions, we compared the available target-point annotations from the five annotators across the 621 events unanimously accepted by all five annotators (Figure~\ref{fig:pseudotarget-reliability}). For each annotator, the leave-one-out consensus was the aggregate target location computed from the other four annotators after withholding that annotator's point. The median distance to this consensus was \(1.93^{\circ}\), and median pairwise disagreement was \(2.39^{\circ}\), well below the median target width of \(13^{\circ}\); median point spread was 5.9\% of the target extent. We regard the approximately \(2^{\circ}\) agreement scale as small enough for a pseudo-calibration target because it spans less than 2\% of the \(110^{\circ}\) camera field of view, is approximately one foveal diameter, and characterizes events accepted by all five annotators. {The \(10.1^{\circ}\) 95th-percentile offset identifies a small tail of larger deviations, while pooling reduced residual correction bias from \(1.89^{\circ}\) for one target to \(0.78^{\circ}\) for eight. This result motivates confidence filtering to exclude high-disagreement targets and consensus-based aggregation of several retained keypoints so that an isolated noisy annotation has less influence on each calibration update.} {As an independent behavioral check, we calculated for each accepted event the minimum angular distance between explicitly calibrated gaze and the human-consensus keypoint within the event window. Across events, this distance had a minimum of \(0.78^{\circ}\), a median of \(1.23^{\circ}\), and a maximum of \(4.26^{\circ}\), confirming that participants' calibrated gaze approached the annotated functional-target centers during the accepted events.}

\section{VLM Based Candidate Validation and Target Localization}\label{sec:vlm-validation}

\textbf{RQ2: Can a VLM use temporal context from egocentric video and images to validate candidate PVA events and localize functional targets that can serve as pseudo-calibration references?} We evaluate this capability across multiple open-source and closed-source VLMs using the same prompting framework and the human-annotated reference labels described in Section~\ref{sec:annotation-protocol}.
\subsection{Candidate-Event Validation and Target Localization}\label{sec:vlm-tasks}

We formulate VLM prompting as a gated two-task framework that separates candidate-event validation from spatial target localization. In \textbf{Task~1}, the high-recall multimodal detector first proposes candidate PVA events based on temporal agreement among head-motion stability, raw-gaze stability, and wearer-hand presence. For each candidate, the prompt presents the VLM with temporally ordered and timestamped egocentric frames captured before, during, and after the detected interval, together with synchronized behavioral measurements and detector metadata. The VLM first describes the ongoing action and identifies the smallest functional feature that may define a PVA target.

The VLM evaluates eight operational criteria spanning task necessity, functional-target precision, temporal relevance, hand-eye consistency, reachability effort, foveal demand, practice effects, and motion stability. For each criterion, it returns a label, confidence, and grounded evidence, then jointly evaluates the complete profile to produce an \emph{accept}, \emph{reject}, or \emph{uncertain} verdict. Sensor measurements provide supporting evidence but cannot override a clear visual or semantic contradiction. Appendix~\ref{app:task1-prompt} gives the full definitions and response format.

\textbf{Task~2} is activated only when Task~1 accepts the candidate. Its prompt receives the validated event summary and target description, the representative keyframe, and up to three preceding frames. The VLM infers the intended action and predicts a tight ROI around the receiving, contact, selection, or alignment feature. The single visual keypoint is defined as the geometric center of that predicted ROI, so the ROI and point are mutually consistent rather than independent predictions. This instruction distinguishes the functional target from visually salient distractors such as the hand, the manipulated object, or the image center.

The ROI corners and center keypoint are represented in normalized coordinates from 0 to 1000 and converted to pixels using the frame dimensions. The structured response also contains localization evidence, confidence, and a visibility label. If the functional target cannot be localized reliably, the model must omit both the ROI and keypoint rather than guess. The complete Task~2 prompt and response schema are provided in Appendix~\ref{app:task2-prompt}.

\subsection{Human Consensus and Evidence for the Four PVA Hypotheses}
\label{sec:hypothesis-validation}

{Five annotators evaluated each candidate event, and unanimous acceptance by all five annotators defined the human-consensus PVA label used to evaluate the VLM. We separated measurements used in candidate-generation from measurements evaluated independently of the detector. Head stability, gaze stability, gaze dispersion, and hand presence are therefore reported descriptively as detector operating characteristics. Inferential comparisons were restricted to measures not used in candidate selection, including gaze-hand distance, fixation duration, pre/post-event motion, and reachability-related measures. To account for temporal dependence in the 10-Hz streams, comparisons were calculated within recordings and tested across recordings; confidence intervals were estimated by resampling recordings 10,000 times, and seven retained primary comparisons were corrected using the Holm procedure. Full procedures are reported in Appendix~\ref{app:statistical-methods}.}

{Evidence for H1 came primarily from human judgments and target geometry: necessity was selected in 90.6\% of accepted judgments, the median target width was \(10.8^{\circ}\), approximately 9.8\% of the \(110^{\circ}\) camera field of view, and 14.2\% of targets were at most \(5^{\circ}\). The median annotator-to-consensus offset was \(1.93^{\circ}\), approximately one foveal diameter, further supporting the spatial-precision component of H1. These results support the criterion that a functional target is relatively small within the complete scene and requires localized central vision. The lower gaze dispersion observed during PVA intervals is reported as an expected operating characteristic of the gaze-stability proposal threshold. H2 was satisfied operationally: Stage~1 required concurrent IMU-based head stability and raw-gaze stability for every proposed event, and the head-gaze stability association was positive in 142 of 143 AEA recordings. This empirically effective stability gate is characterized as a detector operating condition.}

{H3 was supported by evidence independent of hand-presence gating: hand-eye consistency was selected in 77.4\% of accepted judgments, and median gaze-hand distance decreased from \(13.2^{\circ}\) for unanimously rejected events to \(3.2^{\circ}\) for unanimously accepted events. The separate comparison between PVA and control intervals for gaze-hand distance remained significant after correction (\(p_{\mathrm{Holm}}=2.4\times10^{-17}\)). For H4.1, median fixation duration was 0.840~s during PVA intervals and 0.047~s during controls (\(p_{\mathrm{Holm}}=1.9\times10^{-23}\)), event duration increased from 1.4~s to 4.8~s with annotator agreement, and median head angular rate decreased from 0.49~rad/s before the event to 0.22~rad/s during it before returning to 0.52~rad/s afterward (approach versus event: \(p_{\mathrm{Holm}}=8.1\times10^{-24}\)). For H4.2, reachability effort was selected in only 6.6\% of accepted judgments and was not significantly associated with event duration (\(p_{\mathrm{Holm}}=0.633\)). Reachability-demanding events formed only a small subset of AEA and did not provide enough observations for a well-powered comparison across activities and target locations. H4.2 therefore remains inconclusive and requires a larger dataset or a protocol designed to elicit more interactions with difficult-to-reach targets.}

\subsection{Ground Truth and Model Evaluation}
\label{sec:model-evaluation}

For Task~1, unanimous acceptance by all five annotators defined the ground-truth PVA label. For Task~2, consensus human ROIs defined the reference functional target, and the reference keypoint $(x_{\mathrm{gt}},y_{\mathrm{gt}})$ was computed as the ROI center. Models predicted both the ROI and this center keypoint.

We evaluated four closed-source VLMs, Claude Opus~4.6~\cite{anthropic2026claudeopus46}, GPT-4o~\cite{openai2024gpt4o}, Claude~3 Haiku~\cite{anthropic2024claude3haiku}, and Gemini~2.5 Flash-Lite~\cite{google2025gemini25flashlite}, together with five open-source VLMs, Qwen3.5-35B-A3B~\cite{qwen2026qwen35}, Gemma~4 E4B-IT~\cite{google2026gemma4}, Qwen2.5-VL-7B~\cite{bai2025qwen25vl}, Llama~3.2 11B Vision~\cite{meta2024llama32vision}, and Ministral~3 8B~\cite{mistral2025ministral3}. All nine models received the same 1,212 candidates, temporal context, {four-hypothesis prompt}, and output format. Task~1 metrics were accuracy, precision, recall, and macro-F1. Task~2 used intersection-over-union (IoU)-based ROI Average Precision (AP) and normalized ROI-center error. The figure reports AP on correctly accepted Task~1 candidates, so validation and localization must be interpreted jointly. The full evaluation required approximately 36 GPU-hours, or 12 hours using two NVIDIA A100 GPUs and one NVIDIA RTX~5090 GPU concurrently.

\begin{figure*}[t]
    \centering
    \includegraphics[width=\textwidth]
    {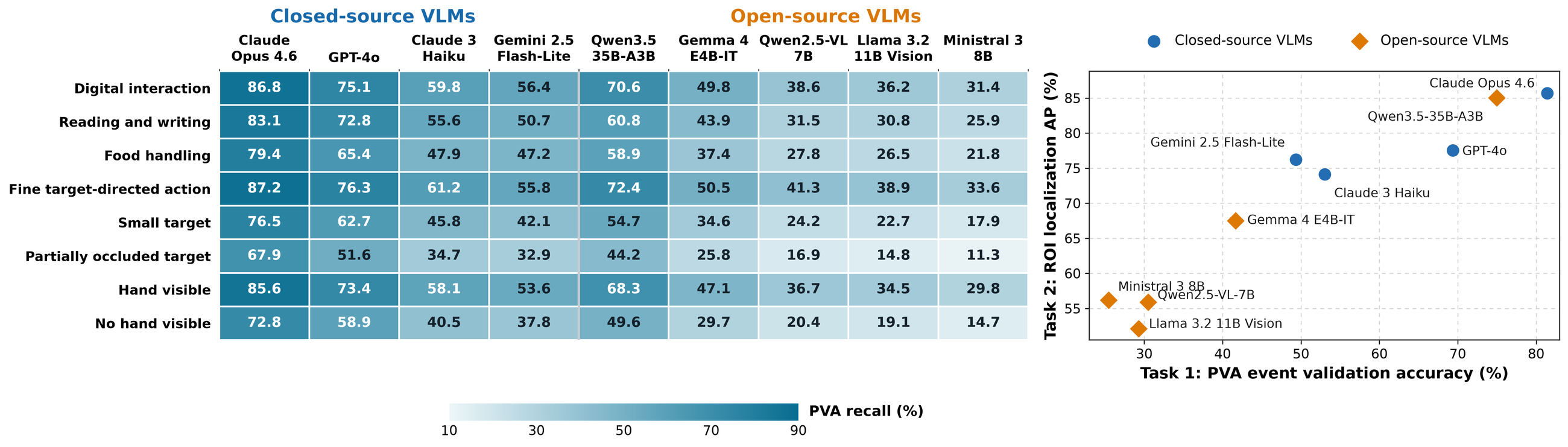}
    \caption{\textbf{Performance of closed-source and open-source VLMs on PVA validation and target localization.}
    \textbf{Left:} Task~1 PVA recall across event conditions; darker cells
    indicate higher recall.
    \textbf{Right:} Task~1 validation accuracy versus Task~2 ROI-localization
    AP. Circles and diamonds represent closed-source and open-source models,
    respectively. Task~2 AP is conditional on correct Task~1 classification.
    }
    \Description{A heatmap of condition-level PVA recall and a scatter plot
    comparing PVA-event validation accuracy with ROI-localization average
    precision across nine vision-language models.}
    \label{fig:vlm-model-evaluation}
\end{figure*}

As shown in Figure~\ref{fig:vlm-model-evaluation}, Claude Opus~4.6 achieved the strongest joint performance, with 81.43\% Task~1 accuracy and 85.68\% Task~2 AP. Among open-source models, Qwen3.5-35B-A3B achieved 74.97\% Task~1 accuracy before fine-tuning on the high-recall candidate set and 85.03\% Task~2 AP. Its localization performance approached Claude Opus~4.6, while event validation remained the larger source of error.

Recall was highest for digital interactions, fine target-directed actions, and events with visible hands, but decreased for small, occluded, and no-hand targets. Qwen3.5-35B-A3B recall decreased from 72.4\% for fine actions to 44.2\% for partially occluded targets and from 68.3\% with a visible hand to 49.6\% without one. We therefore selected it for offline pipeline analysis and parameter-efficient fine-tuning; its open weights also support future deployment without reliance on a commercial API. Occlusion and weak hand-object context remain primary targets for adaptation.

\paragraph{Inference resource profile.}
Inference used a 40,960-token context window with a default 2,048-token generation budget, increased to 24,576 in the ablations. Across the logged configurations, Task~1 used a median of 10,385 prompt tokens and 2,182 response tokens, while Task~2 used 3,001 and 162 tokens, respectively. The 4-bit model occupied approximately 24~GB of device memory. With weights kept resident, median per-call latency was 9.4~s for validation and 2.4~s for localization; when the serving host reloaded the weights for each request, these medians increased to 60.4~s and 46.2~s. These measurements characterize the current offline implementation rather than on-device real-time performance; Appendix~\ref{app:qwen-inference-profile} provides the full token, memory, request-payload, and latency breakdown.

\subsection{Supervised Fine-Tuning}
\label{sec:supervised-fine-tuning}
For the remaining offline pipeline experiments, we fine-tuned Qwen3.5-35B-A3B using SFT with LoRA through LLaMA-Factory~\cite{zheng2024llamafactory,hu2022lora}. Task~1 examples paired temporal context and hypothesis-level evidence with human-consensus labels. Task~2 examples paired validated events with consensus functional-target ROIs and their center keypoints. Data were split by recording to prevent temporal leakage.
{On the held-out test set, fine-tuning increased Task~1 PVA-event validation accuracy from 74.97\% to 89.52\%, an improvement of 14.55 percentage points.} Task~2 ROI-localization AP increased from 85.03\% to 93.98\%, corresponding to an improvement of 8.95 percentage points. Supervision from the human-annotated PVA corpus substantially improved both event validation and spatial target localization while retaining local control over the model weights and inference pipeline.

We configured LoRA with rank $r=16$, scaling factor $\alpha=32$, and dropout rate $0.05$. The vision encoder and pretrained model parameters remained frozen, and only the inserted low rank parameters were optimized. Training used AdamW with a learning-rate of $5\times10^{-5}$, weight decay of $0.01$, a cosine learning-rate schedule, and a warm-up ratio of $0.05$.
Because each training example contained multiple high-resolution frames, the per-device batch size was set to one. Gradient accumulation over eight steps produced an effective batch size of 16 across two NVIDIA A100 GPUs. The model was trained for up to three epochs using bfloat16 precision and gradient checkpointing. We selected the checkpoint with the lowest validation loss and applied early stopping with a patience of one epoch to reduce overfitting.

\section{Implicit Calibration}\label{sec:implicit-calibration}

\subsection{Apparatus and Experimental Protocol}\label{sec:experimental-protocol}

\begin{figure*}[t]
    \centering
    \includegraphics[width=\textwidth]
    {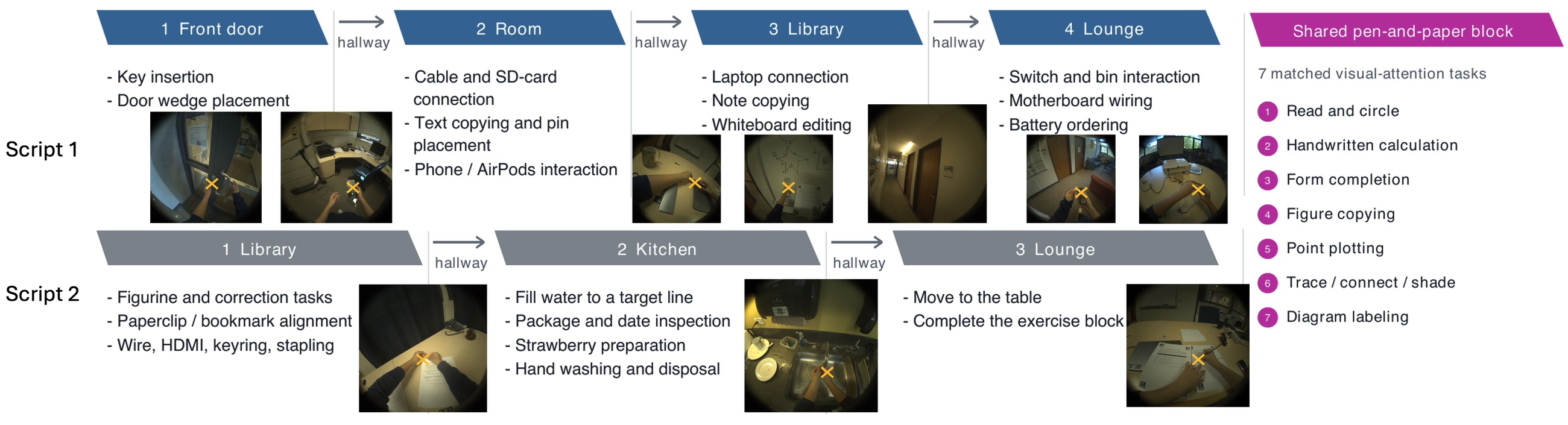}
    \caption{\textbf{Experimental protocol and representative PVA
    events across two scripted routes.}
    Participants completed two continuous multimodal recordings while
    egocentric RGB, eye gaze, IMU, and frontal electroencephalography (EEG) were recorded.
    Script~1 covered four spaces and contained 21 PVA events, whereas
    Script~2 covered three spaces and contained 20 events. Both
    scripts included seven matched pen-and-paper tasks. The representative
    egocentric frames illustrate PVA-relevant activities across locations,
    hallway transitions, and hand-object interactions. Each script was
    performed once under a control condition and once under a disturbance
    condition, during which the glasses were adjusted, participants
    jumped or jogged, and the device was removed and re-worn at predefined
    points.}
    \Description{Diagram of two experimental routes. Script 1 proceeds
    through the front door, a room, the library, and the lounge; Script 2
    proceeds through the library, kitchen, and lounge. Both include a shared
    pen-and-paper block. Eight egocentric images show representative tasks,
    followed by timelines for the control and device-disturbance conditions.}
    \label{fig:experiment-protocol}
\end{figure*}

Figure~\ref{fig:experiment-protocol} summarizes the two scripted routes, their PVA opportunities, and the scheduled device disturbances. {The scripted experiment with 10 participants and the case study with five participants were approved by the University of California San Diego Institutional Review Board (protocol \#190706, ``Neurophysiological dynamics in virtual and augmented reality''), and all participants provided written informed consent before recording. Participants in the scripted experiment were 19 to 26 years old (six men and four women), and none reported a history of vision correction.}

Participants wore Meta Project Aria Gen~1 or Gen~2 research glasses, which recorded egocentric video, eye-tracking data, and inertial measurements. Prefrontal neural activity was recorded using a two-channel CGX PATCH EEG system~\cite{cgx2026patcheeg}, with active electrodes at AFz and Fp2 and Fp1 as the reference. The EEG preprocessing and exploratory event-related potential (ERP) analysis are reported in Appendix~\ref{app:eeg-analysis}. Together, these devices captured the visual, behavioral, and neural signals associated with naturally occurring PVA events.

Participants completed two structured task scripts across laboratory, hallway, library, and kitchen environments. Script~1 contained 14 real-world PVA-eliciting activities, including key insertion, cable and SD card connection, text copying, pin placement, motherboard wiring, and battery arrangement. Script~2 contained 13 activities emphasizing spatially precise functional targets, including bookmark alignment, paperclip placement, wire insertion, keyring insertion, page stapling, and filling a container to a marked level. Each script concluded with seven pen-and-paper tasks involving reading and circling, handwritten calculation, form completion, figure copying, point plotting, shape tracing, and diagram labeling. Scripts~1 and~2 therefore contained 21 and 20 protocol-level PVA opportunities, respectively; these opportunities were subsequently validated rather than assumed to be genuine PVA events. Each recording took approximately 20 to 30 minutes to complete, including participant briefing, device fitting, explicit calibration, walking between locations, and the scripted tasks themselves.

Each script was recorded separately under a no-disturbance control condition and a disturbance-induced condition. In the control condition, participants completed the tasks without deliberately interfering with the glasses. In the disturbance condition, participants adjusted the glasses frame between Batches~3 and~4, performed jumping and jogging between Batches~5 and~6, and removed and re-wore the glasses between Batches~7 and~8. These actions were not counted as PVA opportunities; they introduced device-placement and motion changes for the robustness evaluation.

{Across 10 participants, two scripts, and two conditions, the protocol contained 820 ground-truth PVA opportunities. Stage~1 proposed 1,567 candidates and included all 820 opportunities together with 747 non-ground-truth candidates, yielding 100\% proposal recall and 52.3\% proposal precision. Stage~2 accepted 852 candidates (786 true positives and 66 false positives) and rejected 715 (681 true negatives and 34 false negatives), yielding 92.3\% precision and 95.9\% recall.}

\subsection{Calibration-Frame Selection and Adaptive Calibration}\label{sec:adaptive-calibration}

{All implicit-calibration results in Figure~\ref{fig:calibration-comparison} use events and functional-target ROIs predicted by the fine-tuned VLM.} After a PVA event and its functional target have been validated, we extract fixation-like frames that can be paired with the pseudo-calibration target. Let $\mathbf{G}(t)$ denote the measured angular gaze direction at time $t$. We compute its angular velocity as\begin{equation}v_G(t)=\left\lVert\frac{d\mathbf{G}(t)}{dt}\right\rVert_2 ,\label{eq:gaze-angular-velocity}\end{equation}where $v_G(t)$ is the magnitude of the temporal change in gaze direction. A frame is retained when $v_G(t)$ remains below a predefined velocity threshold and gaze dispersion remains below a dispersion threshold within a short temporal window. This criterion removes frames captured during saccadic transitions while retaining stable fixation-like frames that can be matched to validated pseudo-calibration targets.

Each accepted frame provides a correspondence between the measured gaze location and the center of the validated target ROI. Let $(x_i,y_i)$ denote the normalized horizontal and vertical components of the measured gaze for correspondence $i$, and let $(u_i,v_i)$ denote the normalized center of the corresponding pseudo-calibration target. The required corrections are $\Delta x_i=u_i-x_i$ and $\Delta y_i=v_i-y_i$. We model these corrections as
\begin{align}
\Delta x &= A^{x}x + B^{x}y + C^{x} + N^{x}(x,y), \\
\Delta y &= A^{y}x + B^{y}y + C^{y} + N^{y}(x,y).
\end{align}

The six coefficients $A^{x}$, $B^{x}$, $C^{x}$, $A^{y}$, $B^{y}$, and $C^{y}$ define a user-specific affine correction for translation, scaling, and cross-axis error. The nonlinear components $N^{x}$ and $N^{y}$ model spatially varying residual errors using 25 Gaussian basis functions distributed across the gaze field:\begin{equation}N^{d}(x,y)=\sum_{k=1}^{25}\beta_{k}^{d}\phi_k(x,y),\qquad d\in\{x,y\},\label{eq:nonlinear-correction}\end{equation}where $\phi_k(x,y)$ is the Gaussian basis function indexed by $k$ and $\beta_k^{d}$ is its coefficient for correction dimension $d$. The nonlinear coefficients are learned offline from previous recordings and remain fixed during user-specific online adaptation.

Given $n$ accepted gaze-target correspondences, the design matrix $\mathbf{X}$ contains the measured gaze components, a constant term, and the 25 basis-function values; $\mathbf{Y}$ contains the required horizontal and vertical corrections. The parameter matrix is estimated by\begin{equation}\mathbf{W}^{*}=\underset{\mathbf{W}}{\arg\min}\left[\left\lVert\mathbf{X}\mathbf{W}-\mathbf{Y}\right\rVert_{F}^{2}+\mathcal{R}(\mathbf{W})\right],\label{eq:calibration-objective}\end{equation}where $\mathcal{R}(\mathbf{W})$ regularizes the nonlinear and user-specific parameters. We use regularized iteratively reweighted least squares with a $2.5^{\circ}$ Huber threshold and four reweighting iterations; bandwidth and regularization are selected through recording-stratified five-fold cross-validation.

{Each adaptation batch contains four VLM-validated gaze-target correspondences. During participant evaluation, the nonlinear population correction remains fixed and only the six user-specific affine coefficients are updated. Let $M_i$ denote the user-specific calibration parameter set fitted after incorporating the four correspondences in Batch~$i$. We use a one-step-ahead moving-adaptation protocol: Batch~1 initializes $M_1$; the four samples in Batch~2 are evaluated with $M_1$ before being incorporated to fit $M_2$; Batch~3 is then evaluated with $M_2$, and so forth. Thus, every reported Batch~$b$ result ($b\geq2$) is produced before that batch contributes to the next fit, preventing within-batch leakage.}

\subsection{Comparison with General and Explicit Calibration}\label{sec:calibration-comparison}

{We compare three methods: Meta's released general SocialEye model without in-session user calibration, the explicit calibration supplied by the Meta Project Aria application, and our adaptive implicit-calibration method. SocialEye is an open-source PyTorch model trained on data from more than 1,200 individuals and approximately two million Aria eye-camera frames; its released configuration splits the combined eye-camera image into left and right eye crops, processes them with a ResNet-18 convolutional neural network (CNN) backbone, and regresses gaze yaw and pitch with lower and upper uncertainty bounds~\cite{projectariaeyetracking,projectariaeyetrackingdocs}. This is a standard supported use rather than an artificial ablation: Meta provides general gaze output for every recording with eye-camera data, while personalized output requires in-session calibration; the Project Aria explicit condition therefore provides the corresponding standard calibrated baseline~\cite{projectariaeyegazeformat}. Because the public release does not report held-out angular-error accuracy, the SocialEye errors below are measurements on our protocol rather than a reproduction of a vendor benchmark.} Accuracy is measured as the angular distance, in degrees, between the predicted gaze direction and the validated target direction. {In the implicit-calibration condition, every update uses the center of a VLM-predicted ROI; no human target annotation enters the update.}

\begin{figure*}[t]\centering\includegraphics[width=\textwidth]{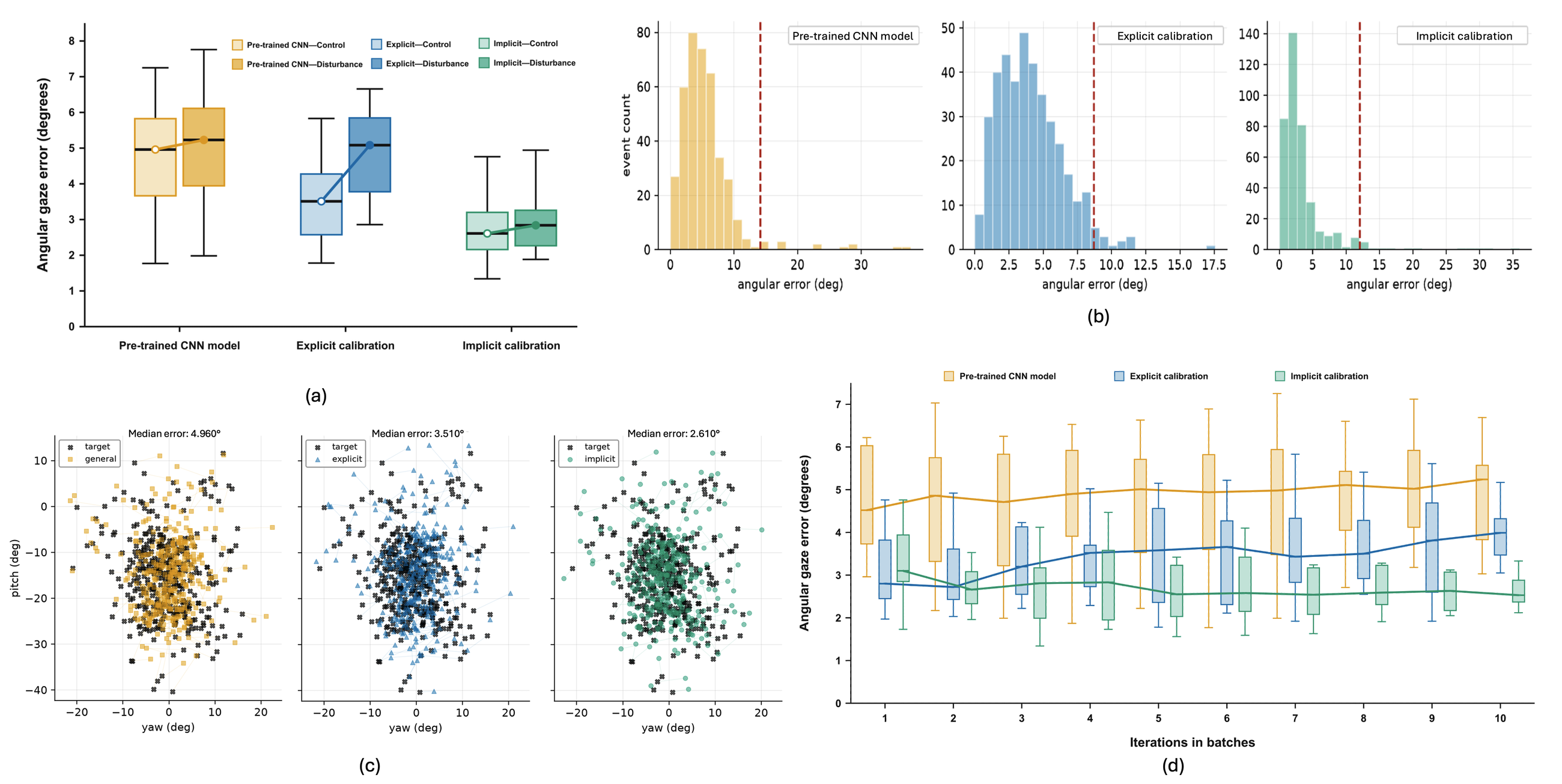}\caption{\textbf{Gaze-estimation accuracy for the general SocialEye CNN, explicit calibration, and implicit calibration.} \textbf{(a)} Angular error distributions under control and disturbance conditions. \textbf{(b)} Error histograms, with red dashed lines indicating method-specific two-standard-deviation thresholds. \textbf{(c)} Target and predicted gaze directions in yaw-pitch coordinates under the control condition. \textbf{(d)} Batch-level angular errors. Batch~1 initializes the implicit model; implicit-calibration results from Batch~2 onward are one-step-ahead evaluations using the model fitted on the preceding batch. {All implicit updates use VLM-predicted ROI centers.}}\Description{Four panels compare three gaze-estimation methods using condition-level boxplots, error histograms, yaw-pitch prediction plots, and batch-level error trends.}\label{fig:calibration-comparison}\end{figure*}

Figure~\ref{fig:calibration-comparison}a compares the three methods under the control and disturbance conditions. Under control, the median angular errors are $4.960^{\circ}$ for the general SocialEye model, $3.510^{\circ}$ for explicit calibration, and $2.610^{\circ}$ for implicit calibration. Under disturbance, the corresponding errors are $5.229^{\circ}$, $5.083^{\circ}$, and $2.839^{\circ}$. Disturbance therefore increases the median error by $0.269^{\circ}$ for the general SocialEye model, $1.573^{\circ}$ for explicit calibration, and only $0.229^{\circ}$ for implicit calibration. Under disturbance, implicit calibration produces a 45.7\% lower median error than the general SocialEye model and a 44.1\% lower error than explicit calibration. At the participant level, an exact two-sided Wilcoxon signed-rank test found a median paired reduction of $2.254^{\circ}$ (95\% participant-bootstrap CI $[1.22^{\circ},3.65^{\circ}]$), $W=1.4$, $p=0.0021$, with a matched-pairs rank-biserial effect size of $r_{\mathrm{rb}}=0.931$; nine of ten participants improved. Figure~\ref{fig:calibration-comparison}b and c further show that implicit calibration concentrates more predictions near their corresponding target directions.

Figure~\ref{fig:calibration-comparison}d shows performance across successive control-condition batches. Batch~1 is used to initialize the implicit model and is excluded from its prospective evaluation. From Batches~2 to 10, each implicit calibration point is evaluated using parameters fitted on the preceding batch, with median errors ranging from $2.53^{\circ}$ to $2.83^{\circ}$. This one-step-ahead protocol shows that the reduction is retained on newly arriving PVA targets rather than measured on the points used to fit the same update.

\subsection{Robustness to Disturbance}\label{sec:disturbance-robustness}

The disturbance sequence covers frame adjustment, intensive head and body motion, and removing and putting on the glasses again. As shown in Figure~\ref{fig:calibration-comparison}a, explicit calibration's median error increases by $1.573^{\circ}$, whereas implicit calibration increases by $0.229^{\circ}$ and retains the lowest overall error. Repeated PVA updates therefore reduce sensitivity to device displacement without requiring another explicit procedure.

\section{Case Study: Personalization under Distribution Shift}
\label{sec:case-study-personalization}

\subsection{Protocol and User Interface}
\label{sec:case-study-interface}

This subsection presents the user-interface design introduced in Section~\ref{sec:tool-design} and evaluates how it supports feedback-assisted personalization.

\begin{figure*}[t]
    \centering
    \includegraphics[width=0.9\textwidth]
    {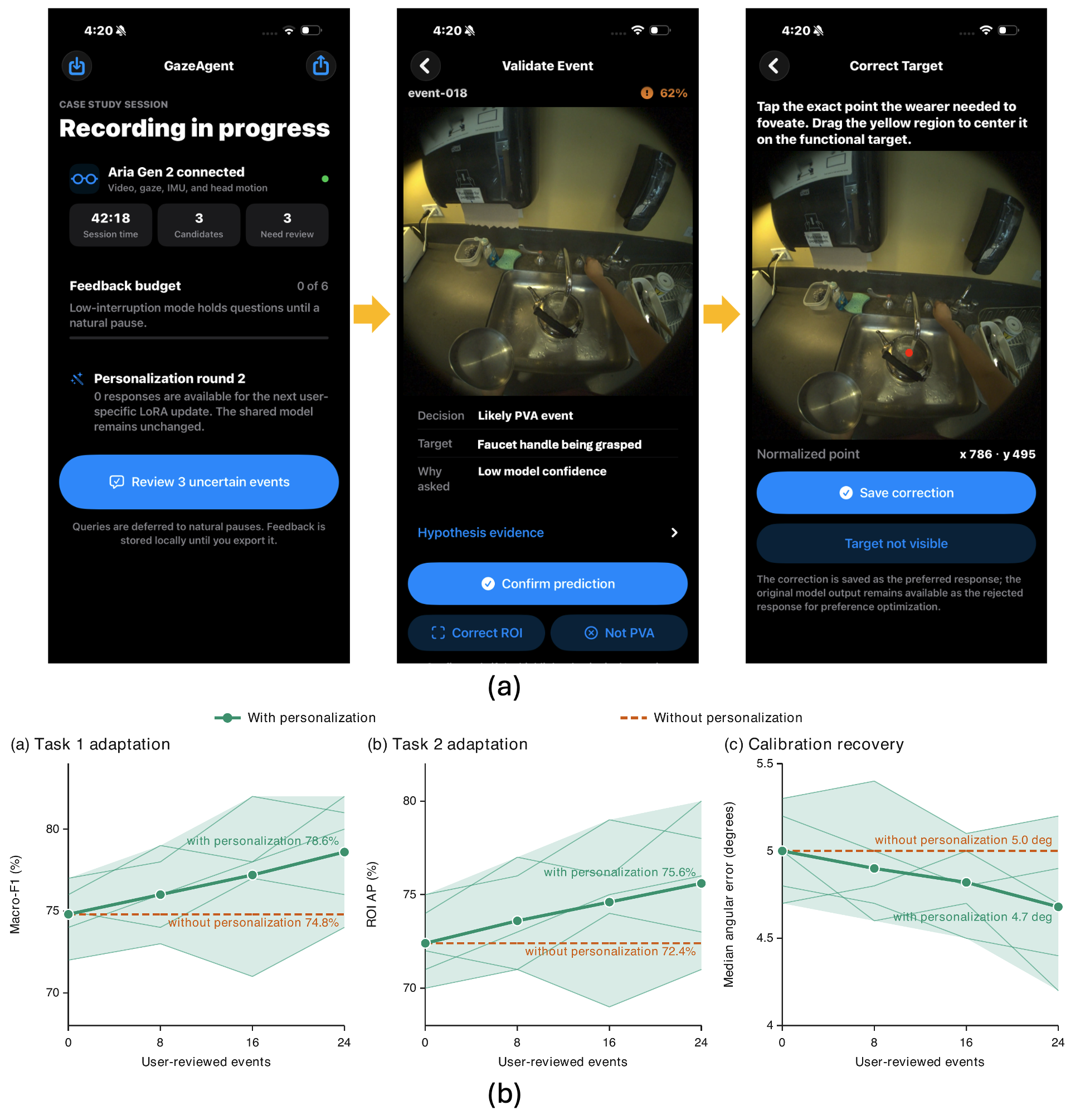}
    \caption{\textbf{GazeAgent interface and feedback-driven personalization.}
    \textbf{(a)} The mobile interface defers low-confidence queries until a natural pause, presents the candidate PVA event and model prediction, and allows the participant to confirm the prediction, reject the event as non-PVA, or correct the functional target by selecting its precise location.
    \textbf{(b)} Changes in Task~1 PVA-event validation, Task~2 ROI-localization, and median angular gaze error as additional user-reviewed events are incorporated into personalization. Thick green curves show the participant mean with personalization, thin green curves show individual participant trajectories, shaded regions indicate the participant range, and dashed orange lines show matched performance without personalization.}
    \Description{The upper panel shows three mobile interface screens for recording a session, reviewing a candidate PVA event, and correcting its target location. The lower panel shows three adaptation plots comparing performance with and without personalization as the number of user-reviewed events increases.}
    \label{fig:gazeagent-personalization}
\end{figure*}

We conducted a case study over two days with five participants to examine optional subject-wise personalization following changes in the wearer or sensing conditions. Each participant completed matched activities with Project Aria explicit calibration, autonomous implicit calibration, and feedback-assisted personalization; condition order was counterbalanced. The study included a stable baseline block followed by a block containing a naturally occurring or experimentally induced disturbance, including frame adjustment, removing and putting on the glasses again, or motion-intensive activity.

For each post-disturbance candidate event, the shared model without personalization and a participant-adapted model generated predictions in parallel. The model without personalization operated in shadow mode and provided a matched reference without affecting the interaction. GazeAgent requested feedback when event-level confidence was low, hypothesis-level evidence was inconsistent, or the predicted ROI varied substantially across adjacent frames. Queries were deferred until natural pauses and constrained by a predefined feedback budget to reduce interruption.

As shown in Figure~\ref{fig:gazeagent-personalization}, the optional interface presented the candidate keyframe, predicted PVA label, target description, confidence, and ROI. Participants could confirm the prediction, reject the candidate, or correct the functional-target ROI; its center was then recomputed as the keypoint. Early post-disturbance feedback was used for personalization, whereas later events were withheld for evaluation.

\subsection{Feedback-Driven Personalization}
\label{sec:case-study-adaptation}

Starting from the globally SFT-adapted Qwen3.5-35B-A3B model, we trained a participant-specific LoRA adapter using DPO~\cite{rafailov2023direct}. For corrected or rejected cases, the human response was the preferred output and the original prediction was the rejected output. Training combined recent feedback with validated examples from a replay buffer, and an update was deployed only when it improved a participant-specific validation buffer.

We compared matched post-disturbance predictions produced with and without personalization. Task~1 was evaluated using accuracy, macro-F1, and PVA recall, whereas Task~2 was evaluated using ROI average precision, mean intersection over union, and normalized target-point error. Downstream calibration was evaluated using angular gaze error. We additionally recorded the number of user-reviewed events, response time, and proportion of predictions corrected by each participant. To examine retention, each personalized adapter was also evaluated on the original AEA test set. Given the small case-study sample, we report descriptive within-participant changes and individual trajectories rather than population-level statistical claims.

{Across five participants, GazeAgent incorporated 24 user-reviewed events per participant, corresponding to 120 feedback events in total. We repeated participant-specific DPO training with three random seeds. The participant-aggregated macro-F1 scores were $91.9\%$, $85.1\%$, and $89.0\%$ ($88.7\% \pm 3.4\%$), compared with $74.8\%$ without personalization. ROI average precision was $84.8\%$, $85.1\%$, and $88.3\%$ ($86.1\% \pm 1.9\%$), compared with $72.4\%$ without personalization. Median angular gaze error was $2.9^{\circ}$, $2.5^{\circ}$, and $2.7^{\circ}$ ($2.7^{\circ} \pm 0.2^{\circ}$), compared with $5.0^{\circ}$ without personalization; $\pm$ denotes the sample standard deviation across seeds.} {The original AEA test set was retained as an external check and changed by an average of $-1.34$ percentage points (range: $-2.4$ to $-0.4$). Figure~\ref{fig:gazeagent-personalization} shows participant-level trajectories across update batches; the validation-gated rule deploys a personalized adapter only after it improves the participant-specific validation buffer.}

\begin{table*}[t]
    \centering
    \caption{\textbf{Within-subject workload, interruption, and preference
    across the three calibration conditions} (\(n=5\)). Quantitative values
    are reported as median [interquartile range, IQR]. Lower
    NASA Task Load Index (NASA-TLX) and interruption scores
    indicate lower burden; bold values mark the most favorable condition.}
    \label{tab:calibration-experience-measures}
    \footnotesize
    \setlength{\tabcolsep}{3pt}
    \renewcommand{\arraystretch}{1.18}
    \begin{tabularx}{\textwidth}{@{}>{\raggedright\arraybackslash}p{0.22\textwidth}YYY@{}}
        \toprule
        \textbf{Measure} &
        \textbf{Aria explicit calibration} &
        \textbf{Autonomous implicit calibration} &
        \textbf{Personalized feedback} \\
        \midrule
        NASA-TLX, median [IQR], 0 to 100 &
        47.5 [46.7 to 55.0] &
        \textbf{41.7 [40.8 to 44.2]} &
        45.8 [43.3 to 48.3] \\
        Interruption rating, median [IQR], 1 to 7 &
        5 [4 to 6] &
        \textbf{1 [1 to 2]} &
        3 [2 to 4] \\
        Preferred for routine use &
        0/5 &
        \textbf{5/5} &
        0/5 \\
        Main perceived benefit &
        Predictable and transparent &
        No explicit action required &
        Allows users to correct errors \\
        Main concern &
        Interrupts the ongoing task &
        Background updates are less visible &
        Queries still require attention \\
        \bottomrule
    \end{tabularx}
\end{table*}

After each condition, participants completed the questionnaire in Appendix~\ref{app:calibration-questionnaire}; Raw NASA-TLX was calculated as the unweighted mean of its six dimensions. As shown in Table~\ref{tab:calibration-experience-measures}, autonomous implicit calibration produced the lowest workload (41.7) and interruption rating (1), compared with explicit calibration (47.5 and 5) and personalized feedback (45.8 and 3). All five participants preferred the autonomous mode for routine use because it required no deliberate calibration action, although they noted that its background updates were less visible; explicit calibration was predictable but disruptive, while personalization enabled error correction at the cost of occasional queries. Given the sample of five participants, these results are descriptive rather than population-level evidence.

\section{Discussion}\label{sec:discussion}
{The results support H1 through task necessity and functional spatial precision, H3 through gaze-hand proximity, and H4.1 through sustained fixation and event duration. H2 is satisfied as an empirical operating property of the detector: Stage~1 admits only intervals with concurrent IMU-based head stability and raw-gaze stability, the head-gaze association was positive in 142 of 143 AEA recordings, and the detector recalled all 820 scripted opportunities. The current data do not support H4.2: reachability effort occurred in only 6.6\% of accepted events and showed no significant association with event duration (\(p_{\mathrm{Holm}}=0.633\)).}

{Because reachability-related events were uncommon, this test was underpowered, and H4.2 remains untested rather than supported. A dedicated protocol should include more reachability-demanding tasks, such as placing a sheet of paper high on a wall, to test whether physical difficulty shifts cognitive effort toward the action target and increases sustained attention. Future work will also test the generality of the empirically effective H2 stability profile on larger, more diverse egocentric datasets such as Ego4D~\cite{grauman2022ego4dworld3000hours}.}

\begin{figure*}[t]
    \centering
    \begin{minipage}[t]{0.32\textwidth}
        \centering
        \includegraphics[width=\linewidth]{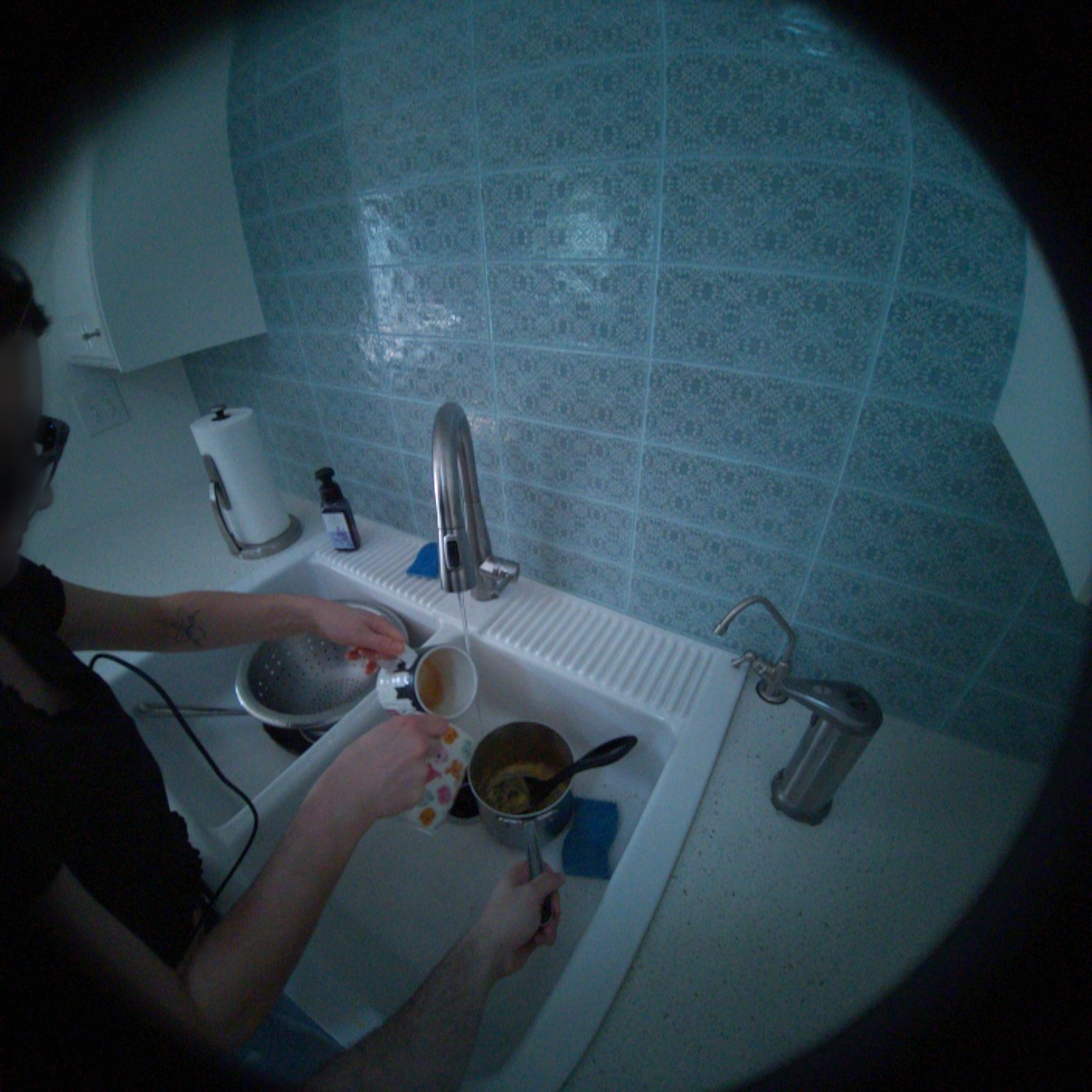}\\[-2pt]
        \textbf{(a) False positive}
    \end{minipage}
    \hfill
    \begin{minipage}[t]{0.32\textwidth}
        \centering
        \includegraphics[width=\linewidth]{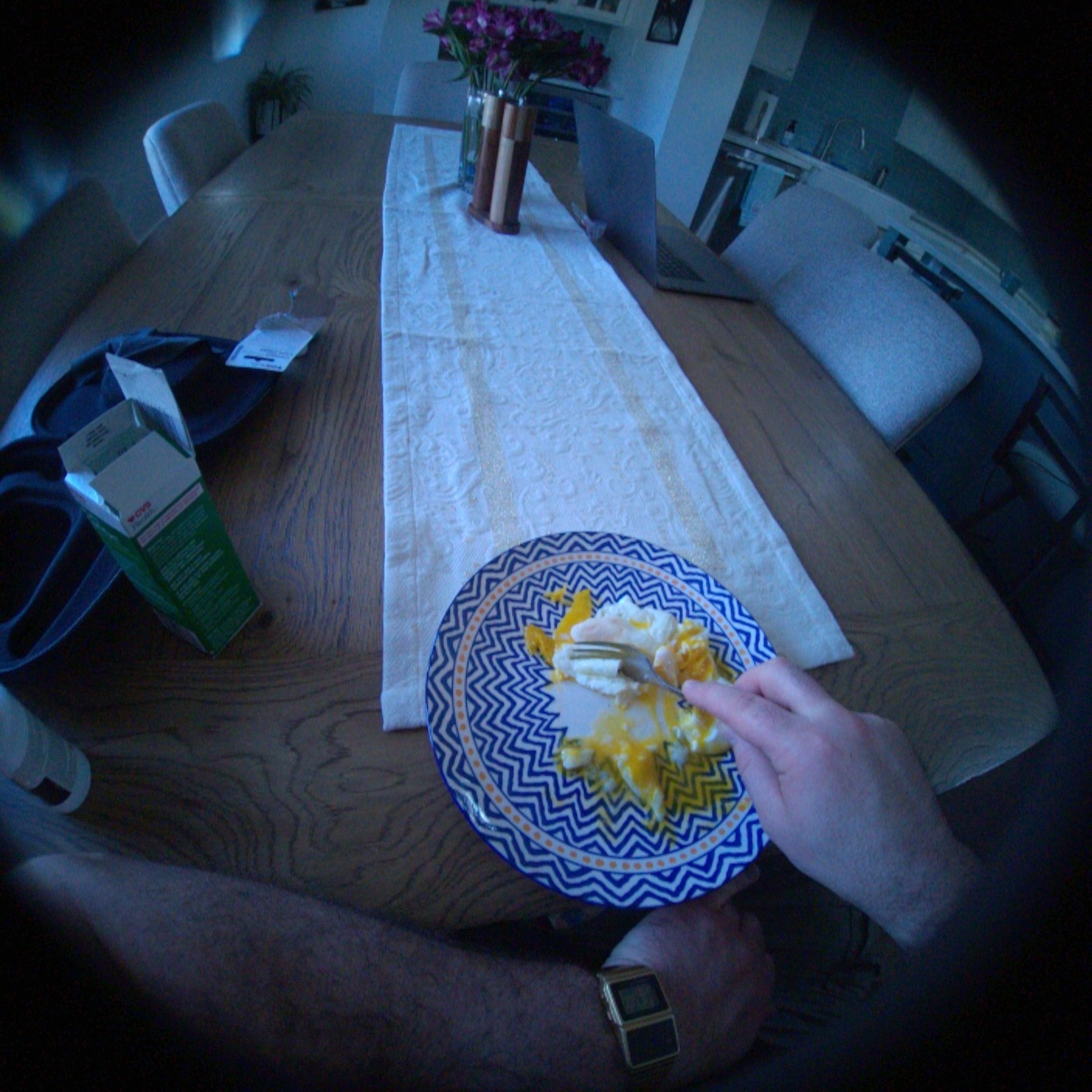}\\[-2pt]
        \textbf{(b) False negative}
    \end{minipage}
    \hfill
    \begin{minipage}[t]{0.32\textwidth}
        \centering
        \includegraphics[width=\linewidth]{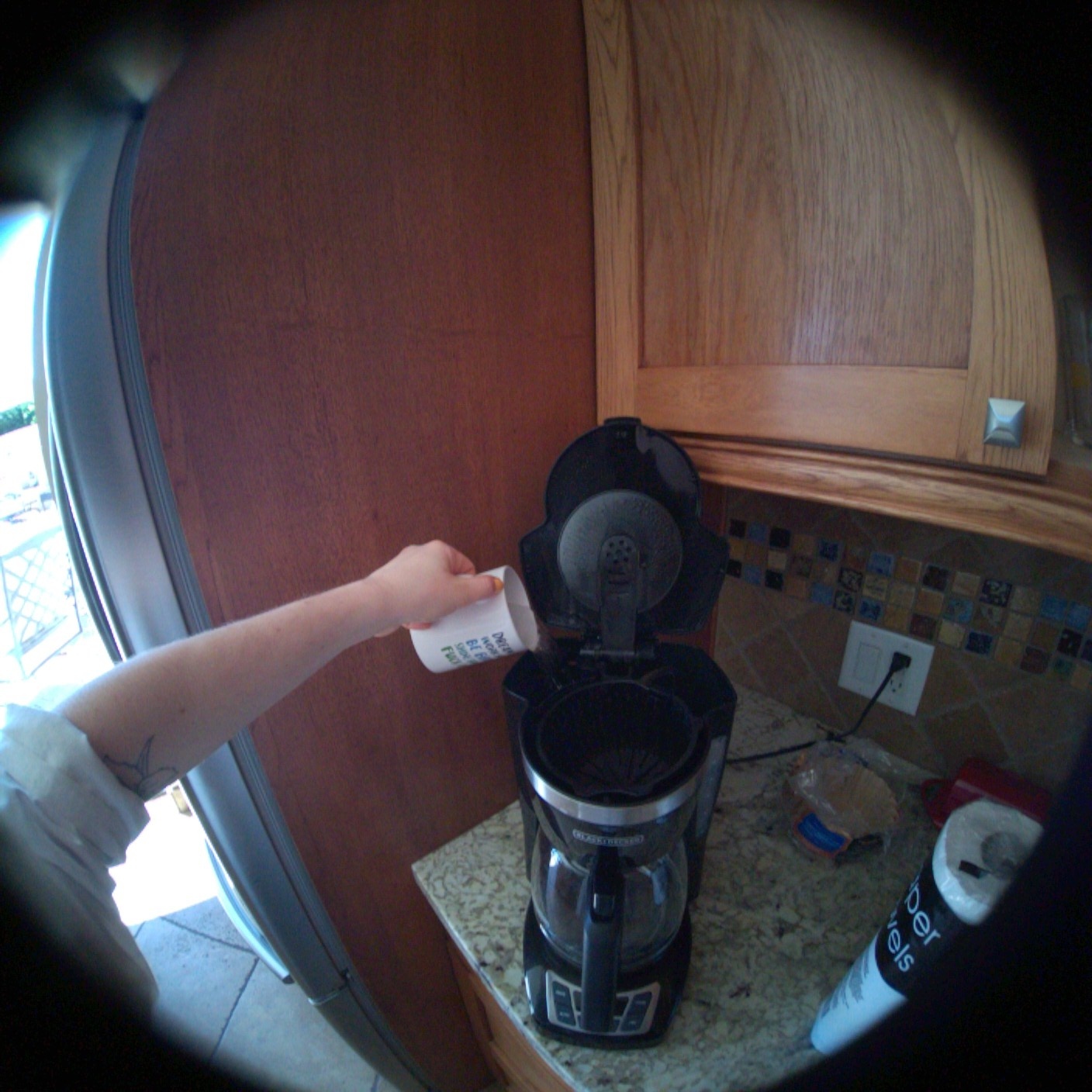}\\[-2pt]
        \textbf{(c) False negative}
    \end{minipage}
    \caption{\textbf{Representative pre-fine-tuning VLM validation errors on AEA.} \textbf{(a)} The model attributes another person's pouring action to the wearer; all five annotators rejected the event. \textbf{(b)} The model treats eating as a routine, coarse action and misses the precise fork-to-food contact point. \textbf{(c)} The model evaluates the entire coffee-maker reservoir rather than its functional opening. All five annotators accepted the events in (b) and (c).}
    \Description{Three egocentric AEA frames illustrating one false-positive and two false-negative PVA validation decisions.}
    \label{fig:pva-errors}
\end{figure*}

We examined pre-fine-tuning VLM errors on AEA events with unanimous five-annotator labels. False positives often occurred when a scene appeared to contain a precise action although the action was not performed by the wearer or the candidate window lacked clear visual attention; Figure~\ref{fig:pva-errors}(a) shows the model attributing another person's pouring action to the wearer. False negatives mainly reflected incorrect target scope: 84\% of their explanations described the target as large, and 64\% invoked a practiced or routine action. Figures~\ref{fig:pva-errors}(b) and (c) show the model overlooking the fork-to-food contact point and the coffee-reservoir opening. These errors primarily concern wearer-hand attribution under H3 and functional-target interpretation under H1. Because a false acceptance can add an incorrect calibration correspondence whereas a false rejection only delays the next update, these patterns motivate stricter actor attribution and functional-target scoping before calibration fitting.

{Stage~1 detector recall cannot be computed on AEA because no exhaustive PVA ground truth existed before candidate generation; human consensus was collected only for the intervals proposed by the detector. Of the 1,212 proposals, 621 (51.2\%) were accepted by all five annotators, yielding 51.2\% proposal precision against unanimous human consensus and showing that semantic validation removed roughly half of the candidate set. This broad proposal pool is consistent with an intentionally permissive, recall-oriented first stage, but the precision value is not a recall estimate because events outside the proposed intervals were not annotated. Future work should independently annotate complete recordings, including detector-negative intervals, to measure detector recall and Stage~2 survival directly.}

{The scripted study provides an exhaustive protocol-level reference: Stage~1 recalled all 820 opportunities, and Stage~2 achieved 92.3\% precision and 95.9\% recall. Stage~2 maintained high recall while its acceptance policy emphasized precision because a false acceptance introduces an incorrect gaze-target correspondence, whereas a false rejection primarily delays the next four-sample calibration update. Stage~2 should therefore be interpreted as precision-oriented semantic filtering that retains 95.9\% of protocol PVA events, together with update availability and time to the next accepted batch.}

{The primary demonstrated advantage is interactional: throughout Stage~1 proposal, Stage~2 validation, and adaptive fitting, calibration evidence is collected automatically from ongoing activity, so no active action is required from the wearer. Together with the disturbance results and the lower reported workload of autonomous implicit calibration, this establishes GazeAgent as a proof-of-concept HCI system for background gaze maintenance. The present contribution is not a claim of real-time edge deployment; rather, it demonstrates the feasibility and user-facing value of the workflow that a smaller or distributed implementation could later operationalize in resource-constrained wearable implementations.}

{The current implementation uses Qwen3.5-35B-A3B for locally controlled offline inference. Even at 4-bit precision, the model occupies approximately 24~GB, which rules out direct execution on current glasses. A practical next step is to use the 35B model and human-verified labels to train a smaller student model, then reduce its cost further through quantization and shorter visual contexts. For scale, a 3B model stored at 4 bits would require approximately 1.5~GB for weights alone, although its total memory use, latency, and accuracy would need to be measured on the target hardware.}

{Until such a model can run on glasses, the workload could be shared with a paired phone. The glasses would detect candidate PVA events and apply the latest calibration, while the phone validates and localizes targets from short event clips; cloud processing could remain an optional fallback when permitted by the user. Because an update is made after four accepted correspondences, this processing can run in the background without blocking gaze output. Our measurements also show that serving configuration matters: keeping the current model resident reduced median validation latency from 60.4 to 9.4~s and localization latency from 46.2 to 2.4~s (Appendix~\ref{app:qwen-inference-profile}). Future work will test whether smaller models preserve target accuracy while meeting the latency, energy, and privacy requirements of everyday use.}

{Because gaze calibration uses egocentric video, it may capture bystanders, screens, or private activities. A deployed system should process only short detector-triggered clips, delete raw frames after ROI and keypoint extraction, and provide a visible recording indicator with simple pause and deletion controls. Future work should examine whether wearers and bystanders understand these safeguards and when consent is needed in shared or private spaces.}

\section{Limitations}\label{sec:limitations}
Task~2 AP is conditional on correct Task~1 predictions, so models can be evaluated on different event subsets and a high localization score does not imply high end-to-end coverage. Future evaluations should additionally report localization on the same human-accepted event set and an end-to-end metric that counts missed PVA events as localization failures. The current study also focuses on AEA and a controlled scripted protocol; broader environments, longer deployments, and more diverse wearers are needed to establish generalizability. Finally, optional personalization queries introduce active effort, whereas the core implicit-calibration results use the autonomous mode in which uncertain candidates are deferred or rejected.

{Earlier implicit calibration systems obtain reference targets from bottom-up saliency or instrumented screen and interface interactions, rather than inferring task-critical functional targets from unconstrained egocentric physical activity. They therefore do not provide an equivalent target-generation pipeline that can be evaluated directly on our Project Aria recordings; a numerical comparison would conflate sensing, task, and target-availability assumptions. We instead evaluate the complete workflow against the uncalibrated SocialEye model and Project Aria's explicit calibration. Future work can adapt a saliency target generator to the same recordings as an input-matched target-source ablation.}

The evaluation does not yet measure how the arrival rate of validated PVA events translates into wall-clock update time, nor does it test controlled contamination by incorrect pseudo-targets. Although the Huber-regularized fitting procedure limits the influence of large residuals, its tolerance to plausible rates of false or mislocalized correspondences remains unverified. Future work should report the number and elapsed time of events required for each update and include corruption-rate ablations that measure when calibration improvement becomes unstable.

\section{Conclusion}\label{sec:conclusion}
We presented GazeAgent, a VLM-guided implicit gaze-calibration pipeline that derives pseudo-targets from naturally occurring PVA events in egocentric AI glasses. GazeAgent combines behavioral sensing with a fine-tuned VLM to detect PVA events and localize functional-target ROIs. In a 10-participant scripted protocol with 820 predefined PVA opportunities, candidate detection achieved 100\% recall and VLM validation achieved 95.9\% recall and 92.3\% precision; implicit calibration achieved $2.84^{\circ}$ median error under device disturbance, 44.1\% lower than explicit calibration. In a five-participant case study, autonomous calibration reduced workload and interruption, and all participants preferred it for routine use. These results establish the potential for autonomous gaze maintenance to support reliable, uninterrupted attention-aware interaction as AI glasses develop into a next-generation general-purpose wearable computing platform.

\bibliographystyle{ACM-Reference-Format}
\bibliography{references}


\appendix
\section{VLM Prompting Framework}
\label{app:vlm-prompts}

This appendix presents the prompts used in our two-task VLM framework.
Task~1 uses image-grounded few-shot prompting to validate candidate PVA
events. Task~2 subsequently localizes the functional
visual target for candidates accepted by Task~1. Event-specific frames,
behavioral signals, detector measurements, and metadata are inserted into the
corresponding prompt at inference-time.

\subsection{Task 1: Candidate-Event Validation}
\label{app:task1-prompt}

\noindent\textbf{System instructions.}
\begin{quote}
\small
You are a vision-language assistant for validating candidate PVA events in
egocentric video. A behavioral detector has proposed
one candidate event using head-motion stability, gaze stability, and
wearer-hand presence. The detector is optimized for high-recall and may
therefore include false positives. Determine whether the given candidate is a
genuine PVA event in which the wearer must look directly
toward a functional target for the action to succeed. Evaluate only the
proposed event; do not identify a different event.

You receive two complementary input modalities:
\begin{enumerate}
    \item \textbf{Egocentric visual context:} temporally ordered and
    timestamped RGB frames. Frames labeled \textit{GREEN} fall within the
    detected interval; \textit{PRE} and \textit{POST} frames provide
    surrounding context; and \textit{KEYFRAME} indicates the representative
    frame.

    \item \textbf{Behavioral signals:} synchronized measurements of head and
    body motion, gaze stability and velocity, fixation behavior, hand presence,
    precision grip, and hand skeleton information.
\end{enumerate}

Treat behavioral signals as supporting evidence. They must not override an
unambiguous visual or semantic contradiction. For example, a high
precision-grip score does not establish a fine manipulation if the frames
clearly show coarse whole-hand carrying.

First, infer the wearer's action and identify the spatially precise functional
feature that the action must reach, contact, insert into, or align with.
Evaluate spatial precision and foveal demand relative to this feature rather
than imposing an absolute size threshold. For example, select an outlet opening
rather than the outlet plate, a keyhole rather than the door, or an on-screen
control rather than the entire device.

For touchscreen interactions, infer hand-eye coupling from the visible
fingertip-to-target action and changes in screen content. A low physical
precision-grip score should not be interpreted as weak hand-eye coupling.

Evaluate the following eight criteria independently:
\begin{itemize}
    \item \textbf{C1: Task necessity.} Would failing to look precisely at the
    target make the action fail or become substantially more error-prone?

    \item \textbf{C2: Functional-target scale.} Is the relevant interaction
    feature relatively small within the full field of view and therefore
    spatially precise, even if the surrounding object is large?

    \item \textbf{C3: Temporal window.} Does the sequence show a meaningful
    fixation-and-action interval rather than a momentary visual occurrence?

    \item \textbf{C4: Hand-eye consistency.} Does the hand action correspond
    spatially and temporally to the inferred visual target?

    \item \textbf{C5: Reachability effort.} Does reaching the target require
    additional positioning or physical effort that increases visual demand?

    \item \textbf{C6: Foveal demand.} Does the action require precise central
    vision rather than broad peripheral awareness?

    \item \textbf{C7: Practice discount.} Could familiarity or motor memory
    reduce the need for visual guidance? Apply this discount only when the
    action can plausibly be completed without precise fixation. Familiarity
    should not discount actions that always require alignment with a small
    feature.

    \item \textbf{C8: Motion stability.} Is the wearer sufficiently stable
    for reliable gaze estimation? Walking, posture transitions, or rapid head
    motion reduce reliability.
\end{itemize}

For every criterion, return a label
(\texttt{pass}, \texttt{fail}, \texttt{uncertain}, or
\texttt{not\_observable}), confidence, visual evidence, sensor evidence,
inference, and its role in the decision
(\texttt{positive}, \texttt{negative}, or \texttt{neutral}).
Do not interpret \texttt{uncertain} or \texttt{not\_observable} as positive
evidence.

Base the final verdict on the complete criterion profile. C1 is necessary but
is not sufficient for acceptance. C2, C4, and C6 constitute the core evidence
for a precise-target event, while C3 and C8 affect temporal validity and
calibration reliability. Reject the candidate if C1 clearly fails or if the
core criteria strongly contradict a PVA interpretation. Return
\texttt{uncertain} when the available evidence is insufficient or conflicting.
\end{quote}

\noindent\textbf{Few-shot demonstrations.}
\begin{quote}
\small
The system instructions are followed by four image-grounded demonstrations
from recordings different from the candidate being evaluated. Each
demonstration is presented as a separate user and assistant exchange containing
a real egocentric keyframe, behavioral evidence, Stage~1 measurements, and a
structured response. Two examples represent valid PVA events, while two
illustrate common false-positive cases.
\end{quote}
\begin{figure*}[t]
    \centering
    \setlength{\tabcolsep}{3pt}
    \begin{tabular}{@{}cc@{}}
        \includegraphics[width=0.47\textwidth]{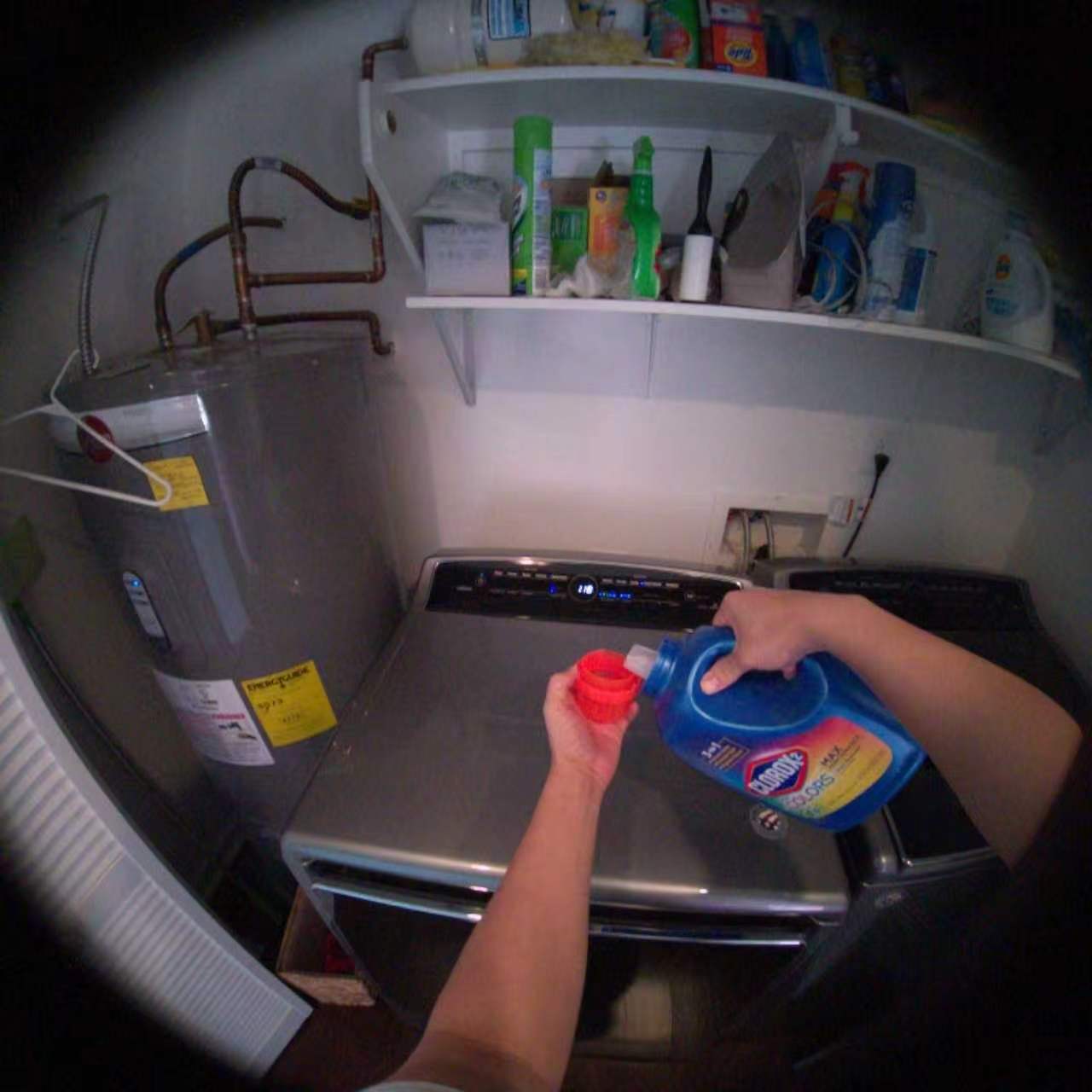} &
        \includegraphics[width=0.47\textwidth]{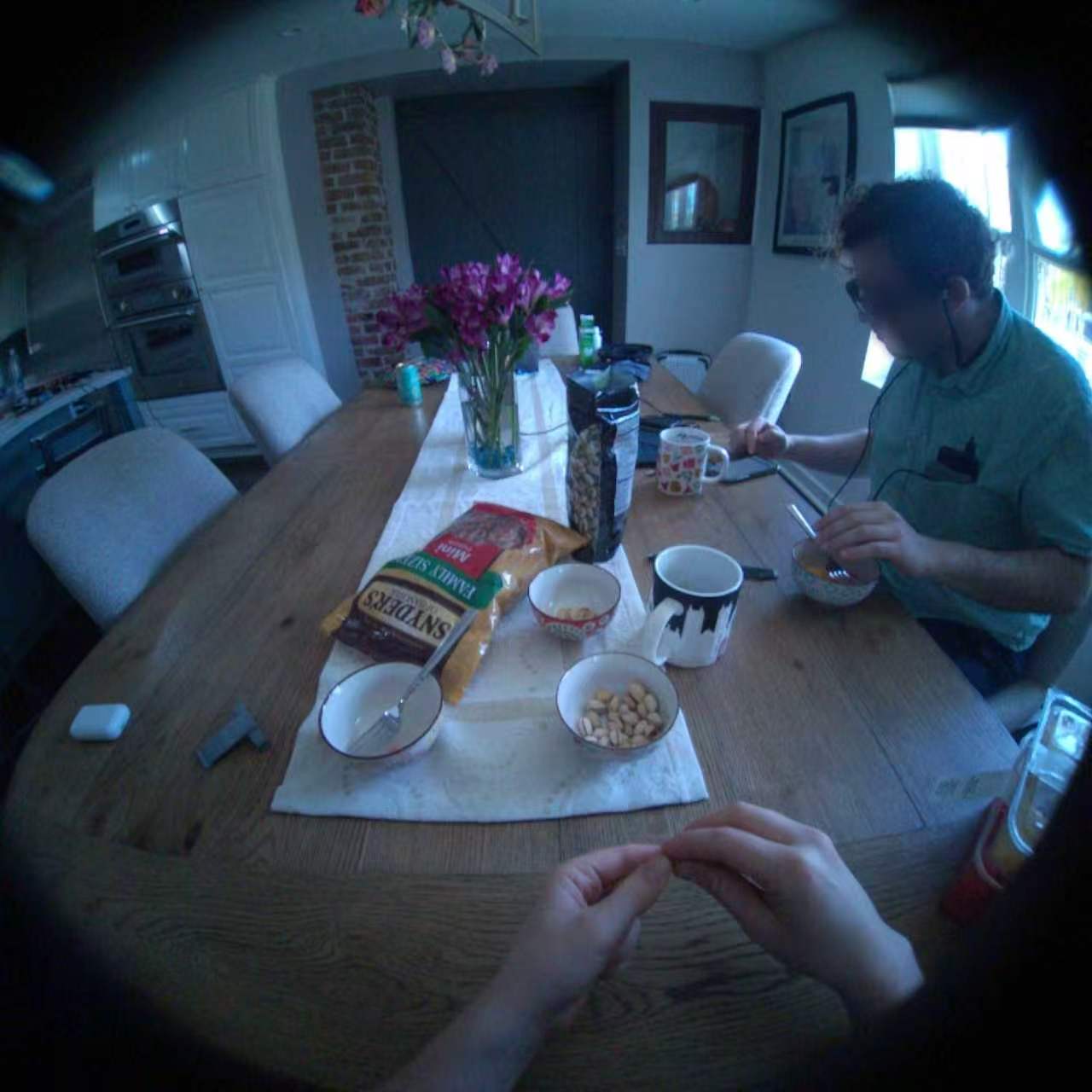} \\
        \small (a) Accepted: detergent pouring &
        \small (b) Accepted: small-item pinch \\[4pt]
        \includegraphics[width=0.47\textwidth]{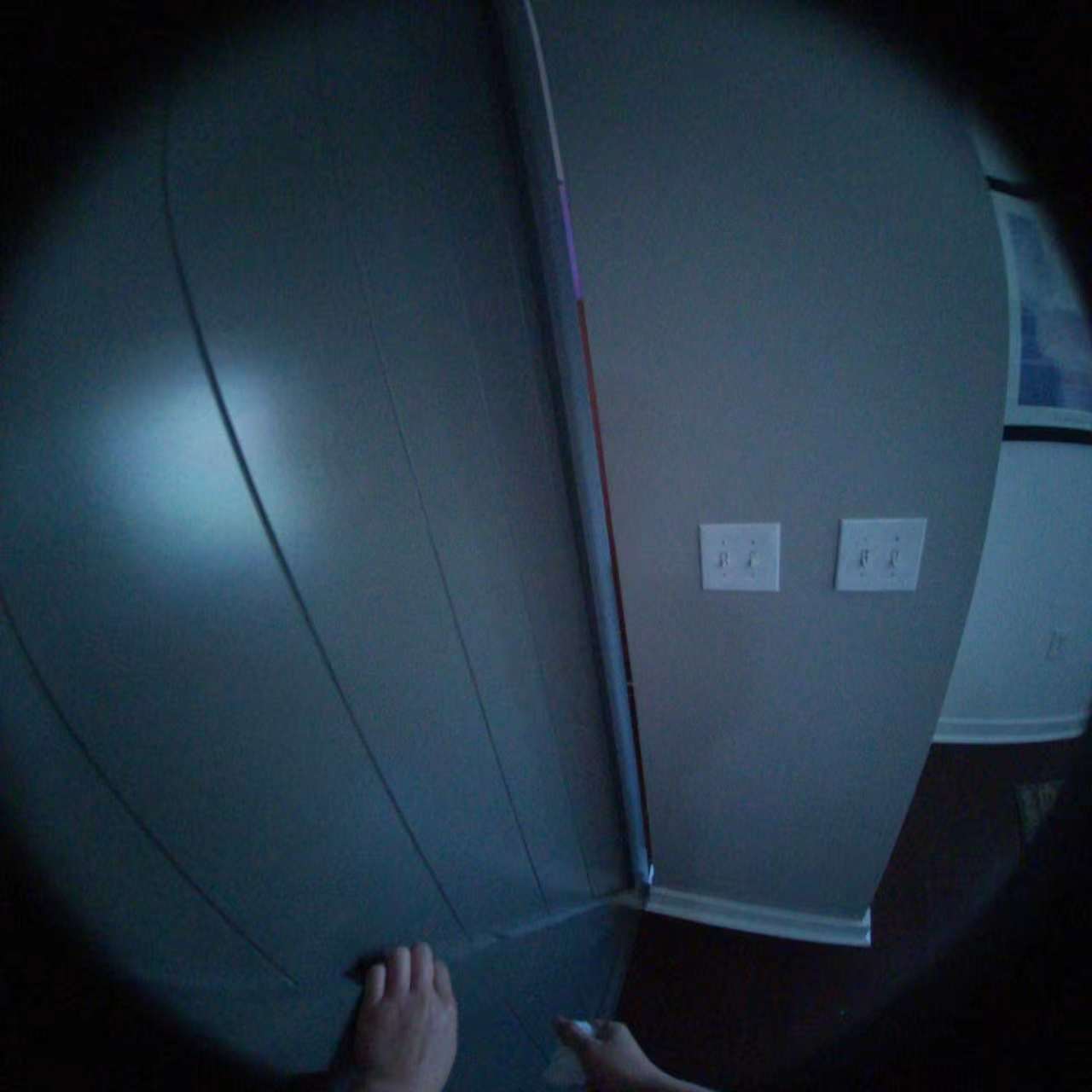} &
        \includegraphics[width=0.47\textwidth]{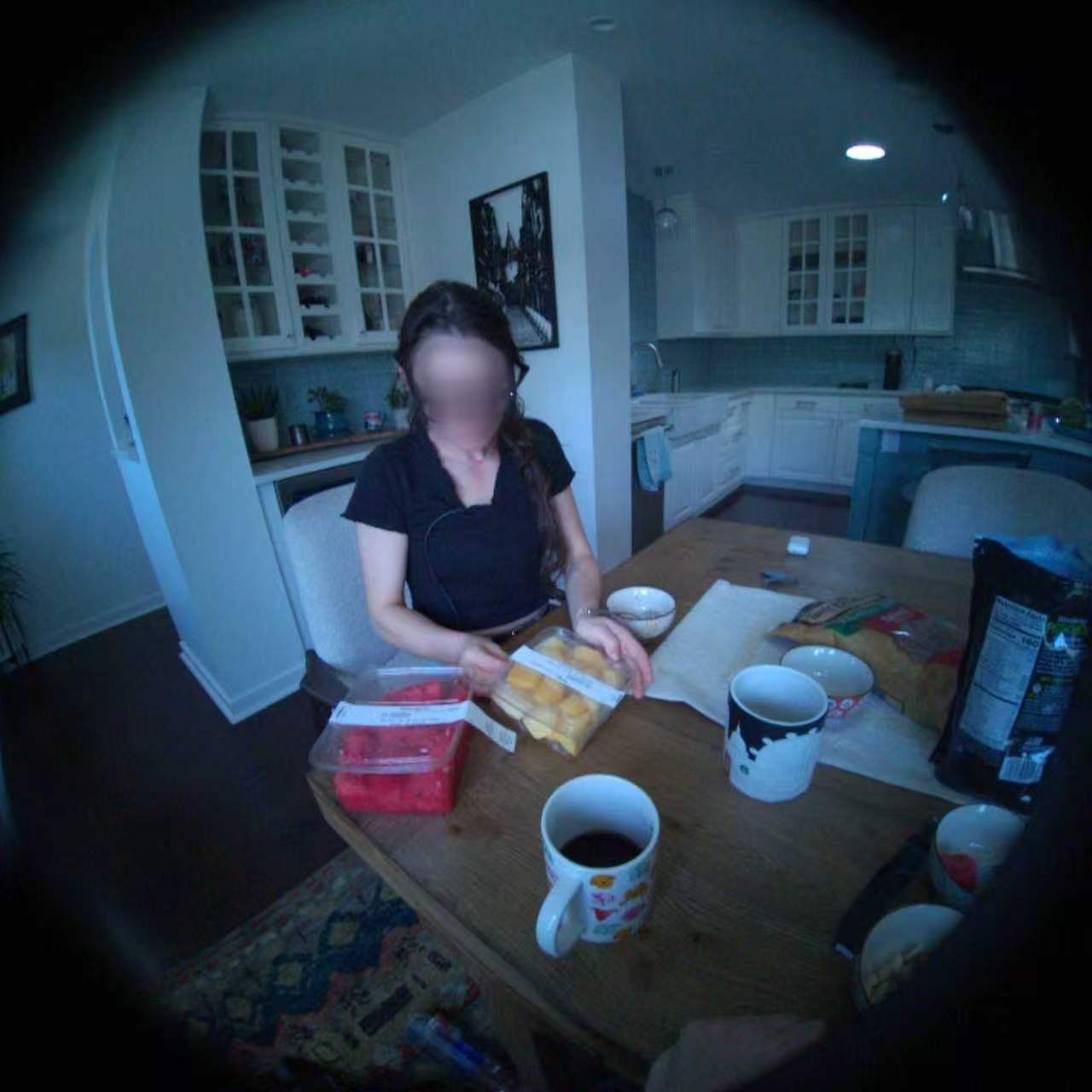} \\
        \small (c) Rejected: flat-panel door push &
        \small (d) Rejected: partner manipulation
    \end{tabular}
    \caption{\textbf{Image-grounded few-shot exemplars used for Task~1 validation.}
    The validation prompt includes two accepted and two rejected examples from recordings different from the candidate under evaluation. The accepted examples require precise alignment with a small functional target: the measuring-cap opening in (a) and the item pinched between the fingertips in (b). The rejected examples show a coarse interaction with a broad door panel in (c) and a manipulation performed by another person rather than the wearer in (d).}
    \Description{Four egocentric few-shot example frames arranged in two rows. The top row contains accepted PVA examples: pouring detergent into a small cap and pinching a small item. The bottom row contains rejected examples: pushing a broad door panel and watching another person manipulate a fruit container.}
    \label{fig:task1-fewshot-examples}
\end{figure*}

\begin{lstlisting}[style=promptbox]
WORKED EXAMPLES

Study the reasoning demonstrated by the following
examples, and then apply the same reasoning process
to the new candidate. These examples come from
different recordings. Do not copy their verdicts,
target descriptions, or evidence.
\end{lstlisting}

\noindent\textbf{Few-shot example 1: Detergent pouring.}
\begin{lstlisting}[style=promptbox]
USER [attached example keyframe]

Observation:
The wearer pours liquid detergent from a large jug
into a small measuring cap held in the other hand.
The liquid stream is visible above the cap opening.

Behavioral signals:
head_stability=0.93; gaze_stability=0.87;
precision_grip=0.22; wearer_hands=2.

Stage 1 measurements:
C3 duration=4.30 s; C4 wearer_hands=2;
C8 head_stability=0.93.

Respond using the Task 1 JSON schema.

ASSISTANT

{
  "decision": "accept",
  "is_precise_visual_attention": true,
  "confidence": 0.92,
  "criteria": {
    "C1": {
      "label": "pass",
      "inference":
        "Visual guidance is necessary to avoid
         spilling or misjudging the fill level.",
      "decision_role": "positive"
    },
    "C2": {
      "label": "pass",
      "inference":
        "The functional target is the small cap
         opening rather than the large jug.",
      "decision_role": "positive"
    },
    "C4": {
      "label": "pass",
      "inference":
        "Both hands coordinate around the same
         small receiving target.",
      "decision_role": "positive"
    },
    "C6": {
      "label": "pass",
      "inference":
        "Aligning the liquid stream with the cap
         opening requires foveal vision.",
      "decision_role": "positive"
    }
  },
  "target_description":
    "the opening of the measuring cap into which
     detergent is being poured",
  "event_summary":
    "The wearer carefully pours detergent into a
     small measuring cap.",
  "rejection_reason": null
}
\end{lstlisting}

\noindent\textbf{Few-shot example 2: Small-item pinch.}
\begin{lstlisting}[style=promptbox]
USER [attached example keyframe]

Observation:
The wearer uses both hands to pinch and manipulate
a small item at the edge of a shared table.

Behavioral signals:
head_stability=1.00; gaze_stability=0.72;
gaze_velocity=21.2 deg/s; precision_grip=0.47.

Stage 1 measurements:
C3 duration=0.80 s; C4 precision_grip=0.62;
C8 head_stability=0.78.

Respond using the Task 1 JSON schema.

ASSISTANT

{
  "decision": "accept",
  "is_precise_visual_attention": true,
  "confidence": 0.85,
  "criteria": {
    "C1": {
      "label": "pass",
      "inference":
        "The two-handed pinch requires visual
         guidance to align the fingertips.",
      "decision_role": "positive"
    },
    "C2": {
      "label": "pass",
      "inference":
        "The manipulated item is a small
         functional target.",
      "decision_role": "positive"
    },
    "C4": {
      "label": "pass",
      "inference":
        "The hands perform a temporally aligned
         fine manipulation.",
      "decision_role": "positive"
    },
    "C6": {
      "label": "pass",
      "inference":
        "The small pinch target requires central
         visual attention.",
      "decision_role": "positive"
    }
  },
  "target_description":
    "the small item pinched between the fingertips",
  "event_summary":
    "The wearer uses both hands to manipulate a
     small item at the table edge.",
  "rejection_reason": null
}
\end{lstlisting}

\noindent\textbf{Few-shot example 3: Flat-panel door push.}
\begin{lstlisting}[style=promptbox]
USER [attached example keyframe]

Observation:
The wearer pushes open a large, featureless door
panel using a flat-palm. No handle or small
mechanism is visible.

Behavioral signals:
head_stability=0.84; gaze_stability=0.96;
gaze_velocity=8.9 deg/s; precision_grip=0.63.

Stage 1 measurements:
C3 duration=0.80 s; C4 precision_grip=0.56;
C8 head_stability=0.72.

Respond using the Task 1 JSON schema.

ASSISTANT

{
  "decision": "reject",
  "is_precise_visual_attention": false,
  "confidence": 0.90,
  "criteria": {
    "C1": {
      "label": "fail",
      "inference":
        "The coarse push does not require precise
         visual guidance.",
      "decision_role": "negative"
    },
    "C2": {
      "label": "fail",
      "inference":
        "The contact area is a broad, featureless
         panel rather than a spatially precise target.",
      "decision_role": "negative"
    },
    "C4": {
      "label": "fail",
      "inference":
        "A flat-palm push is not a fine
         hand-target manipulation.",
      "decision_role": "negative"
    },
    "C6": {
      "label": "fail",
      "inference":
        "The action does not require precise
         foveal alignment.",
      "decision_role": "negative"
    }
  },
  "target_description": "",
  "event_summary":
    "The wearer pushes open a large door panel
     using a flat-palm.",
  "rejection_reason":
    "The action has no spatially precise functional target or
     fine manipulation. High gaze stability and
     precision-grip measurements do not override
     the visual evidence."
}
\end{lstlisting}

\noindent\textbf{Few-shot example 4: Partner manipulation.}
\begin{lstlisting}[style=promptbox]
USER [attached example keyframe]

Observation:
The wearer's partner opens a fruit container
across the table while the wearer observes without
performing the manipulation.

Behavioral signals:
head_stability=0.58; gaze_stability=0.86;
gaze_velocity=12.8 deg/s; precision_grip=0.17.

Stage 1 measurements:
C3 duration=1.80 s; C4 wearer_hands=2;
C8 head_stability=0.80.

Respond using the Task 1 JSON schema.

ASSISTANT

{
  "decision": "reject",
  "is_precise_visual_attention": false,
  "confidence": 0.94,
  "criteria": {
    "C1": {
      "label": "fail",
      "inference":
        "The wearer is a passive observer and has
         no action requiring visual guidance.",
      "decision_role": "negative"
    },
    "C4": {
      "label": "fail",
      "inference":
        "The hands performing the manipulation
         belong to the partner, not the wearer.",
      "decision_role": "negative"
    }
  },
  "target_description": "",
  "event_summary":
    "The partner opens a fruit container while the
     wearer observes.",
  "rejection_reason":
    "The precise manipulation is performed by
     another person. Gaze and hand-tracking signals
     do not establish a PVA event centered on the wearer."
}
\end{lstlisting}

\noindent\textbf{New-candidate reminder.}
\begin{lstlisting}[style=promptbox]
REMINDER: You just studied four worked examples.
The event below is a different, new event,
possibly from a similar-looking activity. Do not
reuse any example's description, coordinates, or
verdict. Compute everything below from the images
and behavioral evidence attached to this message.
\end{lstlisting}

\noindent\textbf{Event-specific candidate input.}
\begin{quote}
\small
\begin{itemize}
    \item Recording: \texttt{<recording\_id>}
    \item Candidate event: \texttt{<event\_id>}
    \item Detected interval: \raggedright from \texttt{<start\_time>} to \texttt{<end\_time>}
    \item Representative frame:\\ \texttt{<keyframe\_id>} at
    \texttt{<timestamp>}
    \item Temporal frames: ordered RGB frames labeled
    \textit{PRE}, \textit{GREEN}, \textit{KEYFRAME}, and \textit{POST}
    \item Behavioral signals: head motion, gaze, and hand measurements
    \item Detector evidence: candidate-generation measurements
\end{itemize}
\end{quote}

\noindent\textbf{Required Task 1 output.}
\begin{lstlisting}[style=promptbox]
{
  "image_id": "<keyframe_id>",
  "event_id": "<event_id>",
  "task":
    "validate_precise_visual_attention_event",
  "decision": "accept | reject | uncertain",
  "is_precise_visual_attention": true,
  "confidence": 0.0,
  "criteria": {
    "C1": {
      "label":
        "pass | fail | uncertain | not_observable",
      "confidence": 0.0,
      "visual_evidence": "",
      "sensor_evidence": "",
      "inference": "",
      "decision_role":
        "positive | negative | neutral"
    },
    "C2": {"same structure as C1"},
    "C3": {"same structure as C1"},
    "C4": {"same structure as C1"},
    "C5": {"same structure as C1"},
    "C6": {"same structure as C1"},
    "C7": {"same structure as C1"},
    "C8": {"same structure as C1"}
  },
  "target_description":
    "<smallest functional target, or empty if
     rejected>",
  "event_summary":
    "<one-sentence describing the action>",
  "rejection_reason":
    "<reason, or null if accepted>"
}
\end{lstlisting}

\subsection{Task 2: ROI/Keypoint Localization}
\label{app:task2-prompt}

\noindent\textbf{System instructions.}
\begin{quote}
\small
You are a vision-language assistant for localizing the functional visual target
of a confirmed PVA event. Task~1 has already established the event as PVA.
Predict a tight region of interest (ROI) around the
functional target and report a single keypoint at the exact geometric center of
that ROI.

The ROI must enclose the spatially precise, semantically meaningful contact,
insertion, selection, or alignment location. Infer the wearer's intended action
from the temporal sequence and localize the intended destination rather than the
hand's current position. Set the keypoint to the ROI center; do not predict a
second, independent point.

Do not select:
\begin{enumerate}
    \item the wearer's hand or fingertip;
    \item the center of the object being held or manipulated when another
    feature receives it;
    \item the center of the image;
    \item a broad object bounding box that does not isolate the functional
    feature; or
    \item a salient or bright region unrelated to the functional target.
\end{enumerate}

Use the validated target description and event summary from Task~1 to constrain
the prediction. Report the ROI corners and center keypoint in normalized
coordinates ranging from 0 to 1000, where $(0,0)$ is the upper-left corner and
$(1000,1000)$ is the lower-right corner of the keyframe. The keypoint must equal
$((x_{\min}+x_{\max})/2,(y_{\min}+y_{\max})/2)$.

If the functional target is occluded, outside the frame, or too blurred to
localize reliably, label its visibility accordingly and omit both the ROI and
keypoint rather than estimating unsupported coordinates.
\end{quote}

\noindent\textbf{Event-specific input.}
\begin{quote}
\small
\begin{itemize}
    \item Recording: \texttt{<recording\_id>}
    \item Event: \texttt{<event\_id>}
    \item Validated target: \texttt{<Task 1 target description>}
    \item Event summary: \texttt{<Task 1 event summary>}
    \item Keyframe: \texttt{<keyframe\_id>} at \texttt{<timestamp>}
    \item Keyframe dimensions: \texttt{<width>} $\times$ \texttt{<height>}
    \item Temporal context: the keyframe and up to three preceding frames
\end{itemize}
\end{quote}

\noindent\textbf{Required Task 2 output.}
\begin{lstlisting}[style=promptbox]
{
  "image_id": "<keyframe_id>",
  "task":
    "predict_precise_visual_attention_roi_and_keypoint",
  "image_width": "<width>",
  "image_height": "<height>",
  "target": {
    "description":
      "<precise functional interaction target>",
    "roi_normalized_0_1000": {
      "x_min": 0,
      "y_min": 0,
      "x_max": 0,
      "y_max": 0
    },
    "keypoint_normalized_0_1000": {
      "x": 0,
      "y": 0
    }
  },
  "confidence": 0.0,
  "evidence":
    "<one-sentence explaining the target location>",
  "visibility":
    "visible | occluded | not_visible"
}
\end{lstlisting}

\noindent
If \texttt{visibility} is \texttt{occluded} or
\texttt{not\_visible}, the model must omit both coordinate fields. For visible
targets, the keypoint is computed from the ROI center and the response is
rejected as malformed if the two outputs disagree.

\subsection{Qwen3.5-35B-A3B Inference Resource Profile}\label{app:qwen-inference-profile}
\paragraph{Context window.}
Inference used a 40,960-token context window and a default 2,048-token generation budget, increased to 24,576 in the ablations. Image content dominated the prompt: a $704\times704$ frame required approximately 261 tokens, or 418 tokens when paired with its frame-level signals, and each event contained a mean of 9.5 frames. The four few-shot demonstrations shown in Figure~\ref{fig:task1-fewshot-examples} contributed 6,103 image tokens and 3,901 text tokens. Across the logged configurations, Task~1 validation prompts had a median length of 10,385 tokens and elicited a median 2,182-token response; the largest prompt contained 31,930 tokens, or 78\% of the context window. Task~2 localization was substantially shorter, with medians of 3,001 input tokens and 162 output tokens.

\paragraph{Memory and request payload.}
The frozen baseline was served at 4-bit precision and occupied approximately 24~GB of resident device memory. The key-value cache was provisioned for the complete 40,960-token window regardless of actual use, compared with a median total sequence length of 12,567 tokens, resulting in 2.4 to 3.3 times more allocation than the logged sequences required. Input frames were downscaled to a 768-pixel long side and base64-encoded at approximately 80~KB per frame, producing a typical request body of approximately 1.4~MB.

\paragraph{Latency.}
When model weights remained resident, median per-call latency was 9.4~s for Task~1 validation and 2.4~s for Task~2 localization. When the serving host reloaded weights for every request, the corresponding medians increased to 60.4~s and 46.2~s. Across 21,468 Task~1 calls, prompt length explained little of the observed latency variance ($R^2=0.013$) and the fitted regression had a 62.8~s intercept, indicating that the fixed per-request loading penalty dominated window length. The complete run consumed 561 endpoint-hours, compared with 64 hours under resident-model execution. These results quantify the current offline implementation and identify persistent model residency, cache sizing, and smaller or distributed models as priorities for future resource-constrained wearable implementations.

\section{Recall-Oriented Candidate Detection Algorithm}
\label{app:candidate-detector}

\begin{algorithm*}[t]
\caption{Recall-Oriented Multimodal Candidate-Event Detection}
\label{alg:candidate-detector}
\begin{algorithmic}[1]
\Require RGB frames \(\mathcal{F}\), IMU stream \(\mathcal{M}\), raw-gaze
stream \(\mathcal{G}\), and recording duration \(T\)
\Ensure Candidate intervals
\(\mathcal{E}=\{(E_e,i_e^{*},Q_{i_e^{*}})\}\)

\State Create a 10-Hz analysis timeline
\(\mathcal{T}=\{t_i\}_{i=1}^{\lfloor 10T\rfloor}\)
\State Align RGB frames, IMU measurements, and gaze observations to
\(\mathcal{T}\)

\ForAll{\(t_i\in\mathcal{T}\)}
    \State Extract a centered 1.0 s IMU window around \(t_i\)
    \State Compute acceleration-magnitude variation and RMS angular velocity
    \State \(S_i^{H}\gets
    \Call{HeadStability}{\text{acceleration variation},
                         \text{RMS angular velocity}}\)

    \State Find the nearest native gaze observation to \(t_i\)
    \If{time difference \(>0.2\) s \textbf{or} angular uncertainty
        \(>15^{\circ}\)}
        \State Mark the gaze observation as invalid
        \State \(S_i^{G}\gets 0\)
    \Else
        \State Extract a centered 0.8 s gaze window around \(t_i\)
        \State Compute gaze angular velocity and angular dispersion
        \State \(S_i^{G}\gets
        \Call{GazeStability}{\text{angular velocity},
                             \text{angular dispersion}}\)
    \EndIf

    \State Detect hands in the RGB frame at \(t_i\)
    \State Classify detected hands as wearer or non-wearer hands
    \State \(N_i\gets\) number of visible wearer hands
    \State \(C_i\gets\) maximum wearer-hand confidence
    \State Store pinch distance, grip aperture, and precision grip descriptors

    \If{\(S_i^{H}\geq0.5\) \textbf{and}
        \(S_i^{G}\geq0.4\) \textbf{and}
        \(N_i\geq1\) \textbf{and}
        \(C_i\geq0.5\)}
        \State \(z_i\gets1\)
        \State \(Q_i\gets
        (S_i^{H}+\epsilon)(S_i^{G}+\epsilon)(C_i+\epsilon)\)
    \Else
        \State \(z_i\gets0\)
        \State \(Q_i\gets0\)
    \EndIf
\EndFor

\State \(\bar{\mathbf{z}}\gets
\Call{MedianFilter}{\mathbf{z},\,0.5\text{ s}}\)
\State \(\bar{\mathbf{z}}\gets
\Call{BridgeGaps}{\bar{\mathbf{z}},\,0.7\text{ s}}\)
\State Remove active runs shorter than \(0.6\) s
\State Form consecutive active runs as preliminary candidate intervals
\State Merge intervals separated by less than \(2.0\) s

\State \(\mathcal{E}\gets\emptyset\)
\ForAll{resulting intervals \(E_e\)}
    \State \(i_e^{*}\gets
    \displaystyle
    \arg\max_{i:\,t_i\in E_e,\,z_i=1} Q_i\)
    \State Add \((E_e,i_e^{*},Q_{i_e^{*}})\) and the component scores
    \(S_{i_e^{*}}^{H}\), \(S_{i_e^{*}}^{G}\), and \(C_{i_e^{*}}\)
    to \(\mathcal{E}\)
\EndFor

\State \Return \(\mathcal{E}\)
\end{algorithmic}
\end{algorithm*}

Each returned \(i_e^{*}\) is considered a \emph{candidate keyframe}, rather
than a confirmed PVA keyframe. The subsequent VLM validation stage determines
whether the interval represents a genuine PVA event, whether the selected
frame captures the moment at which precise attention is required, and whether
its functional visual target can be localized reliably.

\section{Statistical Analysis and Detector Operating Characteristics}
\label{app:statistical-methods}

{Head stability, gaze stability, gaze dispersion, and hand presence contribute directly to candidate selection and are therefore reported as detector operating characteristics. Independent evidence is drawn from human criterion judgments, target geometry, gaze-hand distance, fixation and event duration, pre/post-event motion, and reachability-related measures.}

\begin{figure*}[tbp]
    \centering
    \includegraphics[width=\textwidth]{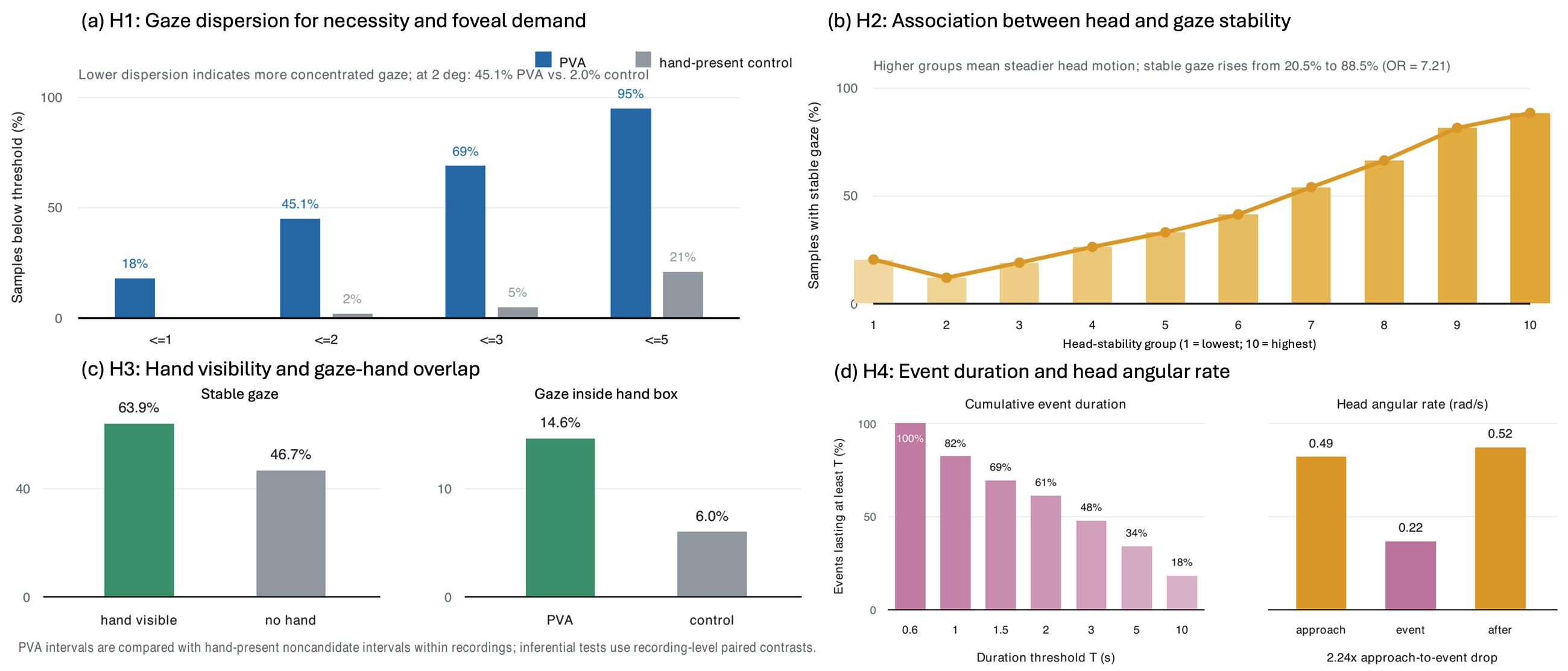}
    \caption{{\textbf{Detector-conditioned operating profiles and temporal context.}
    \textbf{(a)} Gaze dispersion and \textbf{(b)} the head-gaze stability relationship are conditioned by the stability gates in Eq.~\ref{eq:multimodal-agreement}.
    \textbf{(c)} Hand-presence comparisons are likewise conditioned by the visible-hand requirement; the spatial gaze to hand relationship is evaluated independently in Section~\ref{sec:hypothesis-validation}.
    \textbf{(d)} Event duration and pre-, during-, and post-event motion provide temporal context not used to define proposal eligibility.
    Panels (a) to (c) summarize the resulting detector-conditioned profiles.}}
    \Description{Four panels show detector-conditioned gaze concentration, head-gaze stability, hand-presence profiles, and descriptive event-duration and motion context.}
    \label{fig:pva-detector-operating-characteristics}
\end{figure*}

\begin{table*}[t]
\centering
\caption{{\textbf{Independent evidence and detector-conditioned characteristics for the four PVA hypotheses.}
    Inferential (p)-values are reported only for measurements not used in candidate-generation and are corrected using the Holm procedure across the seven retained primary comparisons.}}
\label{tab:hypothesis-statistics}
\scriptsize
\setlength{\tabcolsep}{3.5pt}
\renewcommand{\arraystretch}{1.15}
\begin{tabular}{@{}>{\raggedright\arraybackslash}p{0.06\textwidth}>{\raggedright\arraybackslash}p{0.30\textwidth}>{\raggedright\arraybackslash}p{0.27\textwidth}>{\raggedright\arraybackslash}p{0.31\textwidth}@{}}
\toprule
\textbf{Hypothesis} & \textbf{Independent evidence} & \textbf{Detector-conditioned characteristic} & \textbf{Conclusion} \\
\midrule
\textbf{H1} &
        Necessity was selected in 90.6\% of accepted judgments. Median target width was \(10.8^{\circ}\), approximately 9.8\% of the \(110^{\circ}\) camera field of view, and 14.2\% of targets were at most \(5^{\circ}\). &
        Median gaze dispersion was \(2.19^{\circ}\) for PVA and \(7.47^{\circ}\) for controls, characterizing the stability profile induced by the proposal gate. &
        Task necessity and a functional target that is relatively small within the complete scene are supported. \\
        \addlinespace[3pt]

        \textbf{H2} &
        Stage~1 required concurrent IMU-based head stability and raw-gaze stability for every proposal, and it recalled all 820 scripted opportunities. &
        Stable gaze occurred in 70\% of head-stable samples and 25\% of head-unstable samples; the association was positive in 142 of 143 recordings. &
        H2 is operationally supported: the empirically effective stability gate identifies usable candidate intervals. \\
        \addlinespace[3pt]

        \textbf{H3} &
        C4 was selected in 77.4\% of accepted judgments. Gaze-hand distance decreased from \(13.2^{\circ}\) for unanimously rejected events to \(3.2^{\circ}\) for unanimously accepted events; it was also lower for PVA than control intervals (\(p_{\mathrm{Holm}}=2.4\times10^{-17}\)). &
        Hand-presence is required for proposal generation; stable-gaze comparisons by hand visibility are therefore descriptive. &
        Supported by spatial gaze-hand proximity and human endorsement, with hand presence serving as the proposal gate. \\
        \addlinespace[3pt]

        \textbf{H4.1} &
        Fixation duration was 0.840~s for PVA and 0.047~s for controls (\(p_{\mathrm{Holm}}=1.9\times10^{-23}\)); event duration increased from 1.4~s to 4.8~s with consensus. Head angular rate changed from 0.49 to 0.22 to 0.52~rad/s (approach versus event: \(p_{\mathrm{Holm}}=8.1\times10^{-24}\)). &
        None. &
        Sustained attention and motion settling are supported. \\
        \addlinespace[3pt]

        \textbf{H4.2} &
        Reachability was selected in only 6.6\% of accepted judgments and showed no significant association with event duration (\(p_{\mathrm{Holm}}=0.633\)); these events were uncommon in AEA. &
        None. &
        The current data do not support H4.2; the low event frequency left the test underpowered, and a dedicated reachability protocol is needed. \\
\bottomrule
\end{tabular}
\end{table*}

\paragraph{Behavioral comparisons.}
{Within each recording, inferential comparisons between PVA and control intervals were restricted to measures not used to generate proposals. Hand-present noncandidate intervals were used as controls where applicable to reduce confounding from hand presence. Continuous measures were summarized using medians and binary measures using proportions; recordings were included only when both conditions contained at least 30 samples. The retained primary family comprised precision-grip index, thumb-index pinch distance, gaze-hand distance, the fine-versus-coarse grip proxy, fixation duration, approach-to-event head angular rate, and the association between approach effort and event duration. Paired differences were tested using two-sided Wilcoxon signed-rank tests, 95\% confidence intervals were estimated by resampling recordings 10,000 times, and these seven tests were corrected using the Holm procedure. For applicable independent measures, 200 within-recording circular shifts provided a temporal-alignment negative control.}

\section{Exploratory EEG Analysis of PVA Events}\label{app:eeg-analysis}

PVA events draw on limited perceptual and cognitive resources. Previous studies have shown that visual-task difficulty and working-memory load modulate attention-related ERPs, including the P1 and P300 components, as well as frontal theta-band activity~\cite{pratt2011working,yu2015cognitive,onton2005frontal}. To examine whether PVA events are associated with distinct neural dynamics, we compare validated PVA events with candidate events subsequently rejected as non-PVA. EEG epochs are time-locked to the representative candidate keyframe, which serves as an operational marker of the event. We then compare event-related amplitude, timing, and frontal theta activity between intervals requiring precise foveal guidance and candidate intervals without a validated PVA target. Because P1 is typically strongest over posterior scalp locations, our sparse frontopolar montage is used to examine general event-related and frontal activity rather than to make claims about the spatial origin of individual ERP components.

We analyze signals recorded at AFz and Fp2 using Fp1 as the reference electrode. These electrodes occupy anterior frontal and frontopolar scalp locations in the standardized EEG montage~\cite{seeck2017standardized} and should not be interpreted as direct or spatially specific measurements of the frontal cortex. Nevertheless, activity recorded at frontal sites has been associated with processes including attentional control, working-memory maintenance, planning, and coordination between primary and secondary goals~\cite{koechlin1999role,onton2005frontal}. Because Fp1 serves as the reference, the analyzed channels represent the voltage differences $V_{\mathrm{AFz}}-V_{\mathrm{Fp1}}$ and $V_{\mathrm{Fp2}}-V_{\mathrm{Fp1}}$. We therefore do not use this sparse montage for cortical source localization~\cite{dong2019comparative}.

\paragraph{Synchronization and preprocessing.}
EEG was recorded at 500~Hz from AFz and Fp2 using a two-channel CGX system. Each recording was synchronized independently to the Aria VRS time base by correlating IMU acceleration from both devices to estimate temporal offset and linear clock drift. Alignment was evaluated using acceleration-signal agreement, residual timing error, and support across the full recording. Continuous EEG was processed with a fourth-order, zero-phase Butterworth bandpass filter from 0.5 to 45~Hz and a 60~Hz notch filter. Artifact detection marked unusually large amplitudes, abrupt voltage changes, brief activity bursts, flatlines, and transmission interruptions.

\paragraph{Epoch quality control.}
Epochs were excluded if they overlapped contaminated or interpolated samples, missing samples, data gaps, or two-second safety margins around recording boundaries and transmission interruptions. We also rejected epochs when peak-to-peak amplitude exceeded $100~\mu\mathrm{V}$ or voltage changed by more than $70~\mu\mathrm{V}$ within 20~ms. Baseline standard deviation, absolute linear slope, and detrended root-mean-square amplitude were compared with robust reference distributions pooled across both attention conditions within each recording. When at least 30 candidate epochs were available, this additional criterion rejected an epoch only if at least two measures exceeded a robust $z$-score of 6. Both conditions used identical preprocessing and rejection criteria, without using condition labels or similarity to an expected ERP waveform to determine retention.

\paragraph{ERP analysis.}
The primary analysis examined annotation-locked ERPs for high- and low-visual-attention events in the Script~1 and Script~2 recordings. Each event was linked to the exact timestamp of its RGB frame in the corresponding VRS file and then mapped to the aligned EEG. Epochs extended from 200~ms before to 800~ms after the annotation. Baseline correction subtracted the mean voltage from $-200$ to 0~ms, excluding the event sample. Retained events were analyzed within their labeled condition without pairing events or matching trial counts. For each electrode and condition, epochs were averaged first within recordings and then across recordings with equal weight within each participant. Group waveforms gave equal weight to participants who retained data in both conditions; shading represents the participant mean plus or minus 1.96 standard errors.

\paragraph{Single-trial and spectral analyses.}
Single-trial ERP images display the distribution and timing of retained responses. Each row represents one epoch; the primary images order rows by recording and annotation time and use the same color scale across attention conditions within each electrode. The waveform below each image weights the displayed trials equally, whereas the primary group ERP weights participants equally. Latency-sorted and latency-adjusted images were treated as exploratory supplements. We additionally estimated power spectral density using Welch's method over windows from $-1$ to $+1$~s. These windows passed the same alignment and continuous-signal safety checks but were retained without the additional ERP outlier criteria. Power was averaged in linear units within recordings, across recordings within participants, and across participants before conversion to decibels for display from 1 to 45~Hz. ERP and spectral displays were interpreted descriptively and were not used to establish a statistically significant difference between conditions.

\begin{figure*}[t]
    \centering
    \includegraphics[width=\textwidth]{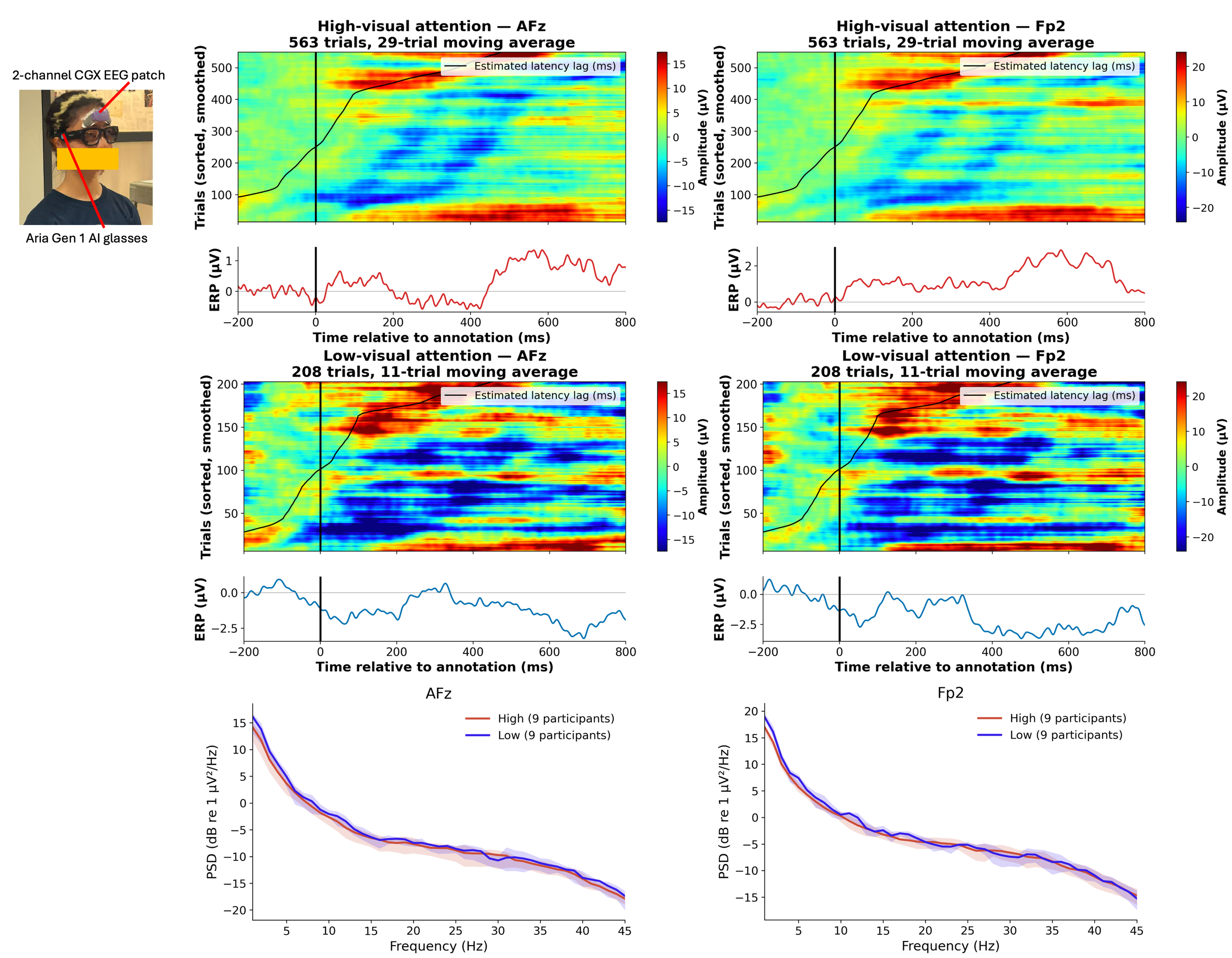}
    \caption{\textbf{Exploratory ERP and PSD analysis of high- and low-visual-attention events.} \textbf{Left:} Wearable recording setup combining Meta Aria Gen~1 AI glasses with a two-channel CGX EEG patch. \textbf{Right:} Trial-sorted, smoothed EEG amplitude maps and averaged ERP waveforms for high-visual-attention PVA events (563 trials) and low-visual-attention rejected candidates (208 trials) at AFz and Fp2. Time zero denotes the annotated event keyframe, and the black curves indicate the estimated latency lag used to order trials. \textbf{Bottom:} Participant-averaged PSD for the high- and low-attention conditions (nine participants each). These frontal measurements characterize event-related activity and are not used for cortical source localization.}
    \Description{A participant wears Meta Aria glasses and a two-channel EEG patch. Four trial-sorted EEG heat maps and averaged ERP traces compare high- and low-visual-attention events at AFz and Fp2. Two PSD plots compare the conditions across nine participants.}
    \label{fig:eeg-erp}
\end{figure*}

\paragraph{Limitations.}
The two-channel montage at AFz and Fp2 cannot support standard multichannel artifact-rejection methods such as independent component analysis or artifact subspace reconstruction. Residual ocular and motion-related contamination may therefore persist, particularly at Fp2. In addition, the annotation timestamp marks the representative event keyframe rather than the physical onset of an attentional change. This temporal uncertainty can broaden an averaged response or reduce its amplitude. The findings should consequently be interpreted as descriptive frontal EEG patterns associated with the annotated attention conditions.

\section{Calibration-Experience Questionnaire}
\label{app:calibration-questionnaire}

\begingroup
\small
Participants completed the following questionnaire immediately after each calibration condition. The six NASA-TLX items used a scale from 0 to 100 in five-point increments, following the official NASA instrument~\cite{nasa2026tlx}; higher values indicate greater workload, with performance anchored by 0 (perfect) and 100 (failure). The calibration-experience items used a seven-point scale from 1 (strongly disagree) to 7 (strongly agree).

\paragraph{NASA-TLX}
\begin{enumerate}
    \setlength{\itemsep}{0pt}
    \setlength{\parsep}{0pt}
    \item How mentally demanding was the activity?
    \item How physically demanding was the activity?
    \item How hurried or rushed did you feel?
    \item How successful were you in completing the activity?
    \item How hard did you have to work to achieve your performance?
    \item How stressed, irritated, or discouraged did you feel?
\end{enumerate}

\paragraph{Calibration experience}
\begin{enumerate}
    \setlength{\itemsep}{0pt}
    \setlength{\parsep}{0pt}
    \item The calibration method interrupted my activity.
    \item The calibration method fit naturally into my activity.
    \item I trusted the gaze estimate produced by the system.
    \item I would be willing to use this calibration method in daily life.
\end{enumerate}

For feedback-assisted personalization, participants additionally rated whether the feedback requests occurred at acceptable moments and whether providing the requested feedback was easy. After all three conditions, participants selected their preferred method for everyday use, briefly explained their choice, and described situations in which they would prefer a different calibration method.
\endgroup

\end{document}